\documentclass[sigplan,nonacm]{acmart}
\usepackage{amsmath}

\usepackage{amssymb}
\usepackage{booktabs}
\usepackage{enumitem}
\usepackage{tabularx}
\usepackage{ragged2e}
\usepackage{tikz}
\usepackage{xspace}
\usepackage{algorithm}
\usepackage{algpseudocode}
\usepackage{listings}
\usepackage{placeins}
\usepackage{needspace}

\lstdefinestyle{tiledramcode}{
  language=Python,
  basicstyle=\ttfamily\tiny,
  keywordstyle=\bfseries,
  showstringspaces=false,
  breaklines=true,
  columns=fullflexible,
  keepspaces=true,
  frame=single,
  xleftmargin=0.5em,
  xrightmargin=0.5em
}

\setcopyright{none}

\newcommand{\fTwo}{\ensuremath{\mathbb{F}_2}\xspace}
\newcolumntype{Y}{>{\RaggedRight\arraybackslash}X}
\newcolumntype{L}[1]{>{\RaggedRight\arraybackslash}p{#1}}
\newcommand{\secref}[1]{Section~\ref{#1}}
\newcommand{\figref}[1]{Figure~\ref{#1}}
\newcommand{\appref}[1]{Appendix~\ref{#1}}
\newcommand{\stepnum}[1]{%
  \tikz[baseline=(char.base)]{%
    \node[
      circle,
      draw,
      inner sep=0pt,
      outer sep=0pt,
      minimum size=1em,
      line width=0.4pt,
      font=\normalfont\normalsize\bfseries
    ] (char) {#1};%
  }%
}

\title{TileDRAM: Compiler-Controlled Tile-to-DRAM Layouts for 3D DRAM GPGPUs}

\begin{document}


\author{Guangyuan Xu}
\affiliation{%
  \institution{ShanghaiTech University}
  \city{Shanghai}
  \country{China}
}
\email{xugy12025@shanghaitech.edu.cn}

\author{Lanxuan Zhang}
\affiliation{%
  \institution{ShanghaiTech University}
  \city{Shanghai}
  \country{China}
}
\email{zhanglx2024@shanghaitech.edu.cn}

\author{Ziyang Li}
\affiliation{%
  \institution{ShanghaiTech University}
  \city{Shanghai}
  \country{China}
}
\email{lizy12024@shanghaitech.edu.cn}

\author{Zhaoshi Li}
\affiliation{%
  \institution{MetaX Integrated Circuits (Shanghai) Co., Ltd.}
  \city{Shanghai}
  \country{China}
}
\email{zhaoshi.li@metax-tech.com}

\author{Yajun Ha}
\affiliation{%
  \institution{ShanghaiTech University}
  \city{Shanghai}
  \country{China}
}
\email{hayj@shanghaitech.edu.cn}

\begin{abstract}
Modern GPU compilers expose rich layout abstractions for mapping tensors to
threads and on-chip resources, but compute-unit-local (CU-local) 3D DRAM architectures
lack a corresponding abstraction for specifying how operand-tile data are arranged
within the DRAM address space.
Without such an abstraction, conventional placement under the existing DRAM
address mapping can subject structured tile accesses to severe
row-buffer conflicts.

We present \textbf{TileDRAM}, a compiler--runtime toolchain that introduces
\emph{tile-to-DRAM layouts} as a compiler-visible abstraction for DRAM placement.
TileDRAM extends the GPU Linear Layout abstraction from compute-side mapping
targets to DRAM placement coordinates and selects layouts per tile access
signature, using recurring tile-local access behavior as the specialization unit.
Since row-buffer locality depends on ordered request
history, TileDRAM evaluates candidate layouts through their row-buffer behavior under
fixed DRAM address mappings. Accesses under the selected layouts are realized through ordinary
address arithmetic, without new address-remapping hardware or ISA support.

In full-trace replay on Accel-Sim/Ramulator2 across 108 MoE FFN cases,
TileDRAM reduces the request-weighted row-conflict rate from 75.7\% to 15.2\%
and achieves a 1.77$\times$ geomean speedup. Retargeting six traces
to six 3D DRAM organizations yields a 2.20$\times$ geomean
speedup under the same compiler abstraction. On an RTX~3090, address transformation adds 0.17\%
geomean overhead over identity.
\end{abstract}

\maketitle

\section{Introduction}
\label{sec:intro}
Memory bandwidth is a first-order system constraint in LLM inference. The
pressure is most acute in feed-forward network (FFN) layers, where dense
matrix projections dominate data movement at small batch sizes. In mixture-of-experts (MoE)
models, routing activates only a few experts per token, yet each activated
expert still performs dense projections over activation and weight tensors
\cite{shazeer2017outrageously,fedus2022switch}. On GPUs, these projections
execute as grouped or batched tiled general matrix multiplications (GEMMs),
whose thread blocks repeatedly fetch operand tiles as they advance along the
reduction (K) dimension. These structured tile streams make high-bandwidth 3D
DRAM an attractive substrate.

\begin{figure}[t]
\centering
\includegraphics[width=0.94\columnwidth]{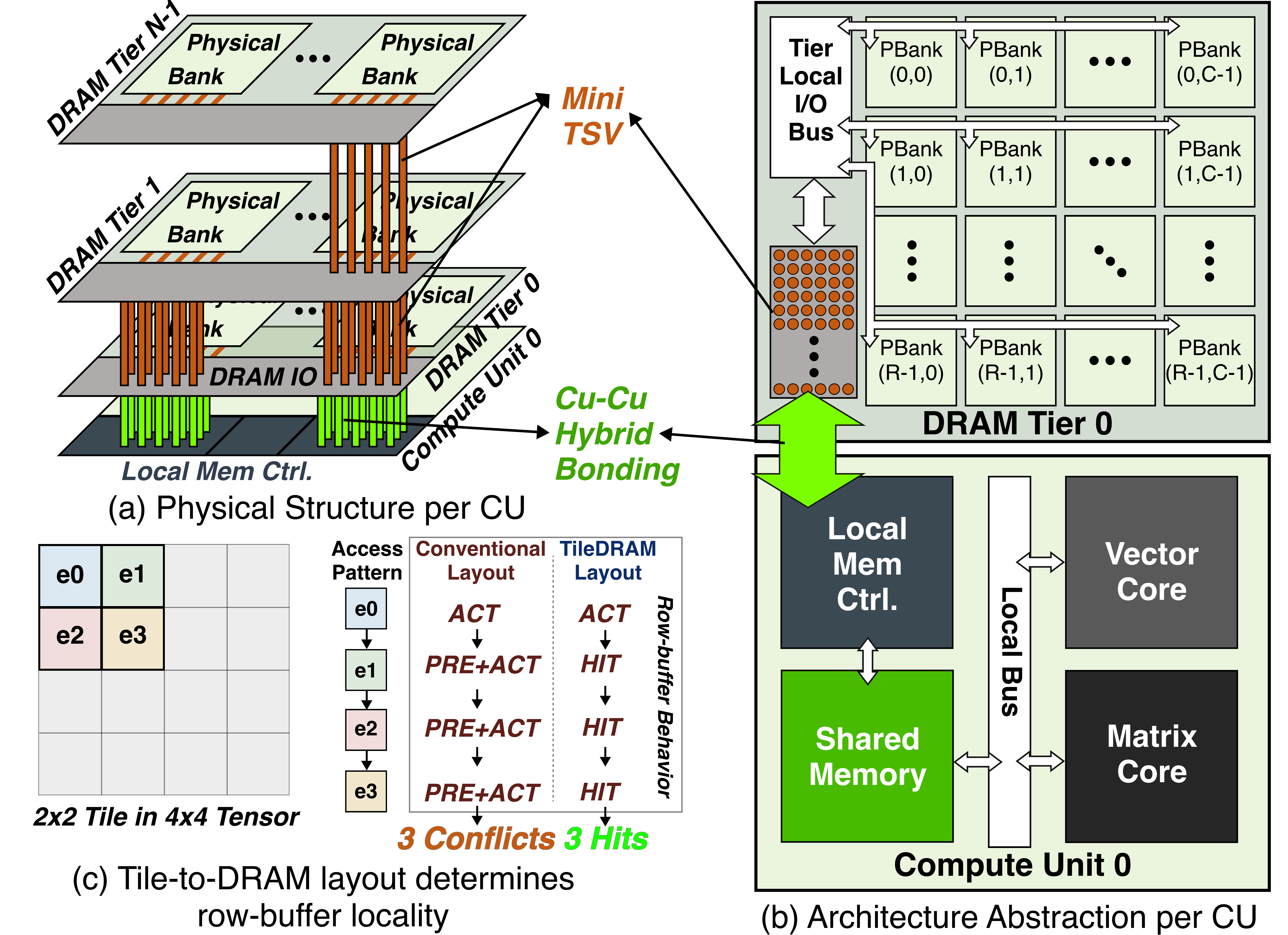}
\caption{\textbf{CU-local 3D DRAM architecture and tile-to-DRAM layout.}
(a)~Hybrid-bonded DRAM stack and logic die.
(b)~One CU--stack pair connected through a local memory controller.
(c)~Row-buffer behavior under the conventional and TileDRAM layouts,
showing how layout choice changes the resulting row-buffer locality.
}
\Description{Three-panel figure. Panel (a) shows a 3D DRAM stack of DRAM tiers
joined by mini-TSVs and hybrid bonding to a logic die that hosts one compute
unit and its local memory controller. Panel (b) abstracts a single CU--stack
pair: the compute unit, with shared memory, vector core and matrix core,
reaches a grid of physical banks in the nearest DRAM tier through a local bus
and its tier-local I/O bus. Panel (c) traces the four elements of a 2x2 tile
inside a 4x4 tensor in access order, and contrasts two placements: the
conventional layout splits them across rows, giving one activation followed by
three precharge-and-activate pairs and thus three row conflicts, while the
TileDRAM layout keeps them in one row, giving one activation followed by three
hits.}
\label{fig:intro-locality}
\end{figure}
Prior 3D DRAM systems motivate two representative chip-wide abstractions for GPGPUs:
using 3D DRAM as the GPU's global memory~\cite{li2025h2llm,pan2025stratum}
or as a GPU-wide last-level cache (LLC)~\cite{loh2012missmap,zhao2019adaptivellc}.
Using the stack as GPU global memory routes its bandwidth through the
package-level endpoint instead of delivering the stack's high local bandwidth
directly, while the LLC abstraction
pays network-on-chip (NoC) area and power to distribute that local bandwidth
GPU-wide~\cite{kahng2009orion2}.
We instead focus on the compute-unit-local
(CU-local) organization motivated in \secref{sec:bg-3ddram-chip},
where each CU accesses its nearby 3D DRAM stack through a software-managed
address-space aperture served by a local memory controller.

However, this high-bandwidth local path also subjects operand
request streams directly to the DRAM controller.
Without warp-level request coalescing and traffic attenuation through the on-chip caches,
numerous fine-grained and potentially redundant memory requests can propagate
to the DRAM subsystem, substantially amplifying contention among concurrent memory requests,
and ultimately reducing the effective memory bandwidth available to the application.
DRAM service efficiency therefore depends on the row-buffer locality of
the resulting row-access sequence.

This dependence on DRAM placement reveals a missing abstraction between compiler-visible
tile semantics and DRAM placement: a \emph{tile-to-DRAM layout} that defines
how operand-tile data are arranged within the DRAM address space.
Under a fixed DRAM address mapping, the tile-to-DRAM layout determines the
physical address stream presented to the memory controller and thereby governs
the row-buffer locality of DRAM accesses (\figref{fig:intro-locality}).

Across our 108 full-trace MoE FFN cases, conventional placement under the
fixed DRAM address mapping yields a 75.7\% request-weighted
row-conflict rate: successive accesses to the same row buffer target different
rows, each paying a precharge and an activation.

Modern GPU software stacks offer rich tile-level abstractions for compute-side
layout and data movement, but provide no corresponding abstraction for DRAM
placement. Triton, CUTLASS, and CuTe provide tiled programming and layout
abstractions
\cite{triton2019,cutlassGemmApi,cecka2026cute}, while cuTile, CuTe DSL,
TileLang, and Gluon expose tiles directly in their programming abstractions
\cite{cuTilePython,cuteDSL,wang2026tilelang,tritonGluon,hagedorn2023graphene}. Linear Layouts
provide a formal representation of distributed tensor layouts as linear maps between labeled
vector spaces over \fTwo \cite{linearlayouts2025}. Across these systems, the
operand tile is already an established compiler-visible unit for specifying
layout and data movement. However, these abstractions stop at the compute side:
they determine how tile data map to threads, registers, and shared memory, but
not how operand-tile data are arranged within the DRAM address space. DRAM
placement therefore remains outside the compiler's tile-level layout
abstraction.

Prior work improves row-buffer locality by modifying the DRAM address mapping
in the memory controller
\cite{liu2018valley,rotaru2025xordrammapping,zhang2022sdam}. Application-specific
synthesis can optimize one global mapping for a representative workload trace
\cite{rotaru2025xordrammapping}, but that mapping must therefore accommodate
the heterogeneous access
patterns within the trace (\secref{sec:bg-stable-specialization}). Independently
specializing these patterns at the controller side would require maintaining
multiple mapping contexts and selecting the appropriate context for each
request \cite{zhang2022sdam}.

Taken together, these observations yield two key insights.
First, CU-local integration provides high local bandwidth, but compiler-awareness
of DRAM placement is necessary to fully utilize the bandwidth. Placement should
therefore become an explicit compiler-controlled decision.
Second, we identify recurring tile-local access behavior as a stable unit
for specializing placement.
This granularity also aligns naturally with existing GPU
software abstractions, where operand tiles are already compiler-visible units for
layout and data movement.

Guided by these insights, we propose \textbf{TileDRAM}, a compiler--runtime
toolchain that introduces tile-to-DRAM layouts as a compiler-visible
abstraction for compiler-controlled DRAM placement. TileDRAM represents this
abstraction as \emph{tile-to-DRAM Linear Layouts}: reusable \fTwo-linear maps
that extend the Linear Layout abstraction from compute-side resources to DRAM
placement coordinates. Each layout maps logical operand-tile coordinates to
DRAM placement coordinates.

Although access behavior varies with operand properties and tile
configurations, the same ordered tile-local access behavior recurs across many
tile instances. TileDRAM captures this with a \textit{tile access signature}
that combines tile configuration, operand properties, and the ordered
tile-local access sequence.
The tile-to-DRAM mapping itself is stateless,
with each tile coordinate mapped without reference to access history.
However, the induced row-buffer behavior is stateful, because whether a request
hits an open row depends on preceding requests in the ordered tile-local access sequence.
TileDRAM therefore evaluates candidate layouts by simulating row-buffer state
transitions under each signature's ordered access, selecting the layout with
the fewest row conflicts.

An offline solver selects reusable layouts for each tile access signature;
the compiler applies them during address generation, and the runtime supplies each
tile's placement without new address-remapping hardware or ISA support.
Retargeting to
different 3D DRAM organizations reselects layouts while reusing the same
abstraction and compilation path.

Using NVBit traces of Triton/Gluon kernels \cite{triton2019,tritonGluon,villa2019nvbit} replayed with
Accel-Sim/Ramulator2 \cite{khairy2020accelsim,luo2023ramulator2}, TileDRAM
reduces the request-weighted row-conflict rate from 75.7\% to 15.2\%
and achieves a 1.77$\times$ geomean kernel
speedup across 108 MoE FFN cases. Across six full-trace streams
retargeted to six 3D DRAM organizations, TileDRAM reduces row
conflicts by 79.2\% and achieves a 2.20$\times$ geomean speedup.
On an NVIDIA GeForce RTX~3090, address transformation adds 0.17\% geomean
overhead over identity across the 108 measured kernel configurations.
Together, these results establish compiler-controlled DRAM placement as a new
optimization dimension: with the DRAM address mapping left unchanged, the
compiler-selected layout alone accounts for the gain.

Our main contributions can be summarized as follows:

\begin{itemize}[leftmargin=*,topsep=0.1em,itemsep=0.05em,parsep=0pt]
\item We introduce \emph{tile-to-DRAM layouts} as a compiler-visible
abstraction for DRAM placement, and represent them as \emph{tile-to-DRAM Linear
Layouts}: \fTwo-linear maps that extend GPU Linear Layouts from compute-side
resources to DRAM placement coordinates, with tile access signatures as their
stable specialization unit.

\item We characterize the row-buffer locality impact of conventional DRAM
placement in CU-local 3D DRAM GPGPUs. Across 108 full-trace MoE FFN cases,
conventional placement yields a 75.7\%
request-weighted row-conflict rate, showing that high local bandwidth
alone does not ensure efficient DRAM service.

\item We present \textbf{TileDRAM}, a compiler--runtime toolchain that selects a
layout per tile access signature, requiring neither
address-remapping hardware nor ISA support.

\item We evaluate TileDRAM using a full-trace framework integrating Triton,
Accel-Sim, and Ramulator2 across 108 MoE FFN cases plus a six-stream,
six-organization retargeting sweep.
TileDRAM consistently reduces row conflicts, improves performance,
and incurs only 0.17\% geomean address-transformation overhead on an RTX~3090.

\end{itemize}

\section{Background and Motivation}
\label{sec:background}

\subsection{3D DRAM and the CU-local Architecture}
\label{sec:bg-3ddram-chip}

Recent 3D DRAM architectures place multi-tier DRAM structures close to
logic through dense vertical integration. In the CU-local organization modeled
here, the vertically integrated DRAM structure is connected to a separate logic
die through face-to-face Cu--Cu hybrid bonding
\cite{li2025h2llm,niu2022hbbonding,pan2025stratum}. The logic die can host memory
controllers and near-memory compute in recent designs
\cite{li2025h2llm,nair2026raptor,pan2025stratum}. This stacked integration
provides a high-bandwidth local DRAM--logic path, while a distinct
package-level endpoint connects the GPU or xPU to the stack
\cite{li2025h2llm,nair2026raptor,pan2025stratum}.

This separation makes the package-level endpoint the bandwidth boundary for
chip-wide access. Stratum reports 19.01--30.34~TB/s of internal bandwidth
across its memory tiers \cite{pan2025stratum}. Exporting a stack through an
HBM-style package interface instead offers roughly 0.8--1.2~TB/s per stack in
HBM3/HBM3E \cite{jedecHbm3,skhynixHbm3e}, making the package-level endpoint
16--38$\times$ narrower than the modeled internal path. This gap motivates
placing compute near the stack rather than exporting its bandwidth
\cite{li2025h2llm,pan2025stratum}. Recent industrial inference hardware
provides a concrete instance of this CU-local principle. OpenAI's Jalape\~{n}o
adopts a memory-sliced architecture in which each compute core slice has a
local view of its own HBM4 slice and the chip omits a global last-level cache,
avoiding contention on a unified global memory subsystem
\cite{ho2026jalapeno}. Remote and global memory accesses use a separate
general-purpose interconnect.

A transparent GPU-wide LLC offers a contrasting global abstraction: it makes
the stack a shared last-level cache reachable from every CU. Requests and
responses therefore traverse a chip-wide NoC and arbitration path. The
associated routers, buffers, and links consume area and power even when the
target cache line resides in that stack
\cite{kahng2009orion2,zhao2019adaptivellc}. More generally, GPU-global-memory
and GPU-wide-LLC organizations place package-level or chip-wide interconnects
on the common path to stack-resident data. A CU-local organization instead
keeps common accesses within a local compute--memory domain, while retaining a
separate path for remote or global communication.

Motivated by this locality-oriented organization, we model the CU-local design
shown in \figref{fig:intro-locality}(b), where each CU accesses its nearby 3D
DRAM stack through a local memory controller, and analyze one CU--stack pair at
a time. TileDRAM's architectural boundary is a software-managed CU-local
aperture in a dedicated device-physical address space. Requests to this
aperture reach the local controller without virtual-address translation,
warp-level request coalescing, or caching; all other addresses traverse the
conventional memory hierarchy. The local stack remains a conventional banked,
row-buffered DRAM system \cite{mutlu2007stfm,rixner2000memory}.

\subsection{MoE FFNs Present Structured Operand-Tile Streams}
\label{sec:bg-moe-operand-tile-streams}

\begin{figure}[!b]
\centering
\includegraphics[width=\columnwidth]{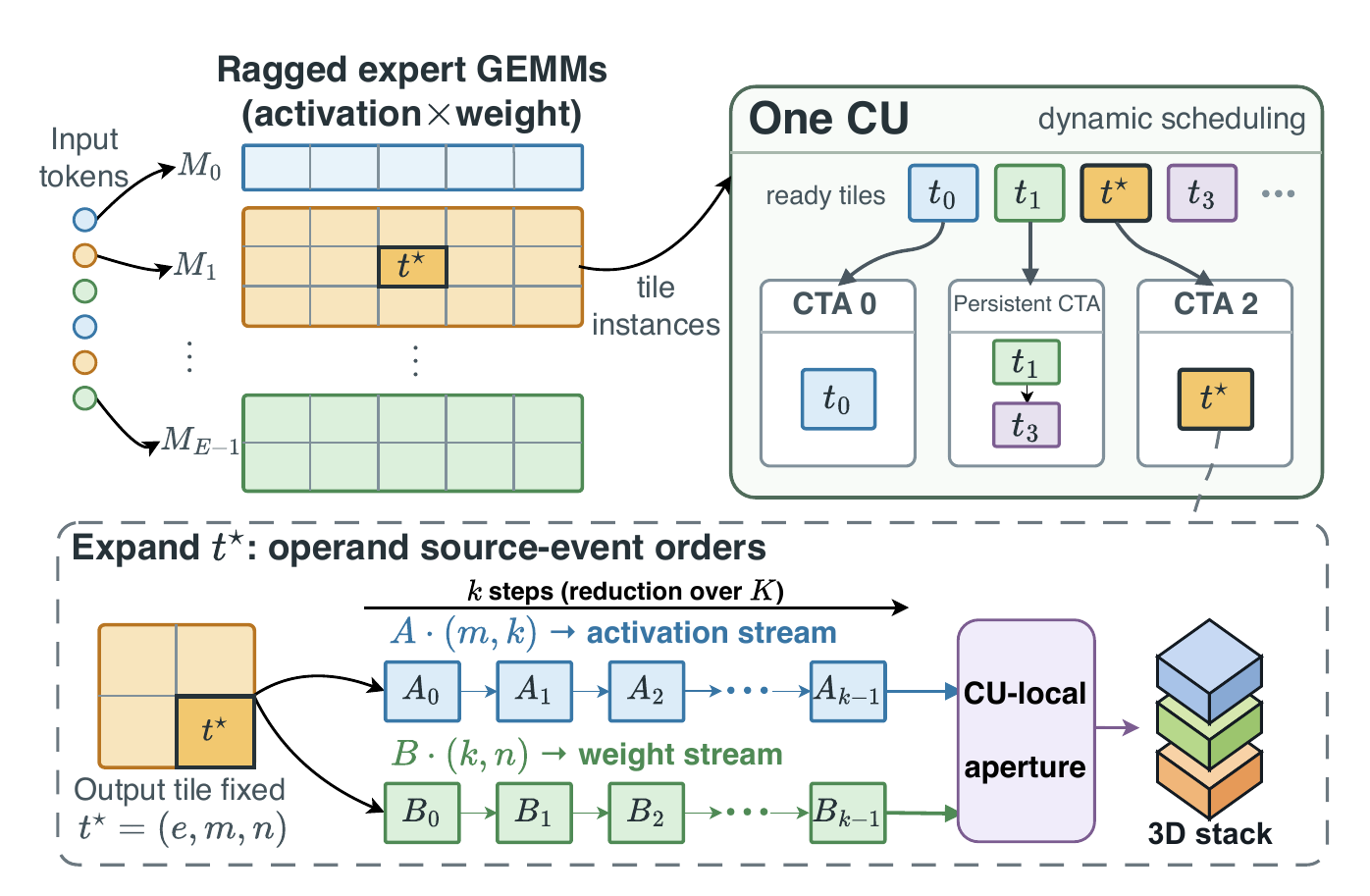}
\caption{Routing leaves ragged per-expert row counts $M_e$; grouped GEMM turns
each $(m,n)$ output tile into a CTA work unit, and dynamic scheduling places
tile instances on resident CTAs of one CU (a persistent CTA may consume
several). Expanding $t^*$ shows the separate ordered tile-local access sequences
for the activation and weight operands presented at the CU-local aperture.}
\Description{Panel (a) traces routed tokens through router/dispatch into
unequal expert groups marked by their ragged $M_e$ dimension, binds a compact
 grouped-GEMM descriptor, and aligns the resulting $(m,n)$ output tiles in a
 thread-block grid. The grid dynamically assigns tile instances to resident
 CTA blocks on one CU; the selected orange tile $t^*$ is shown in a persistent
 CTA sequence with grid-stride assignment, then expanded into panel (b). Panel (b) shows the same tile's repeated
$k$-step progression feeding separate activation and weight event rails; both
rails preserve their local order and terminate independently at the CU-local
aperture beside the paired 3D DRAM stack.}
\label{fig:bg-moe-cu-local}
\end{figure}

MoE FFNs provide a representative workload for studying DRAM placement
within this CU-local boundary: their inference traffic is structured before it
becomes a DRAM address stream. Routing activates a small set of expert MLPs for
each token, and each activated expert
applies dense matrix projections over activation and expert-weight tensors
\cite{shazeer2017outrageously,fedus2022switch}. After routing, the active experts
can be represented as a group of dense GEMM
problems with runtime per-expert row counts:
\[
  C^{(e)}_{mn} = \sum_k A^{(e)}_{mk} B^{(e)}_{kn},
  \qquad 0 \le m < M_e,\ e \in \mathcal{E}.
\]
Here, $e$ identifies an active expert, $M_e$ its routed-token count, $n$ the
output-feature coordinate, and $k$ the hidden/reduction coordinate. Each problem
retains its own $M_e$, the ragged expert dimension left by routing and
dispatch \cite{tritonGroupedGemmTutorial,cutlassGroupedScheduler}.

Grouped GEMM implementations tile each problem's $(m,n)$ output space into
cooperative-thread-array (CTA) work units and advance through $k$ by
repeatedly loading activation and
expert-weight tiles \cite{cudaProgrammingGuide,cutlassGemmApi}. Each CTA executes
on one CU at a time and generates, for each operand stream, an ordered
tile-local access sequence of copy events. Dynamic scheduling assigns tile instances
to CTAs, and persistent CTAs may process multiple instances. Across the grouped
workload, these CTA instances jointly cover each expert problem and may reside
on multiple CUs; \figref{fig:bg-moe-cu-local} follows one representative tile instance and its CTA
residency on one CU. The resulting kernel presents structured,
ordered tile-local access sequences at the CU-local boundary defined above
(\figref{fig:bg-moe-cu-local}). Runtime scheduling determines how CTAs resident
on a CU interleave; this cross-tile interleaving remains separate from each
tile's local access pattern.

Under the standard roofline model
\cite{williams2009roofline,nvidiaDLPerfBackground}, one 2-byte touch of each
FP16/BF16 input and output gives the ideal per-expert GEMM intensity:
\begin{equation}
  I_{\mathrm{ideal}} = \frac{MKN}{MK+KN+MN}\ \mathrm{FLOP/B},
  \label{eq:bg-gemm-intensity}
\end{equation}
taking $M=M_e$ for one expert problem. For an FFN projection with $K=4096$ and
$N=14336$, for example, $M=64$--$128$ gives
$I_{\mathrm{ideal}}=62.7$--$123.1$~FLOP/B; our evaluated suite
(\secref{sec:eval-questions-setup}) spans $55.4$--$123.0$~FLOP/B.

Using published dense-compute and external-memory specifications for conventional
GPUs \cite{nvidiaA100Datasheet,nvidiaAdaArchitecture,nvidiaH100TensorCoreGpu},
the corresponding GDDR/HBM ridge points are 153--295~FLOP/B.
Stratum's reported 19.01--30.34~TB/s aggregate in-stack bandwidth
\cite{pan2025stratum} and the H100 whole-GPU peak
\cite{nvidiaH100TensorCoreGpu} together give a local-memory ridge reference of
32.6--52.0~FLOP/B. This ratio is unchanged under proportional partitioning of
compute throughput and local-memory bandwidth. The evaluated ideal-intensity
band therefore lies below conventional GPU ridges but above this local-memory
reference.

\begin{figure}[!htbp]
\centering
\includegraphics[width=\columnwidth]{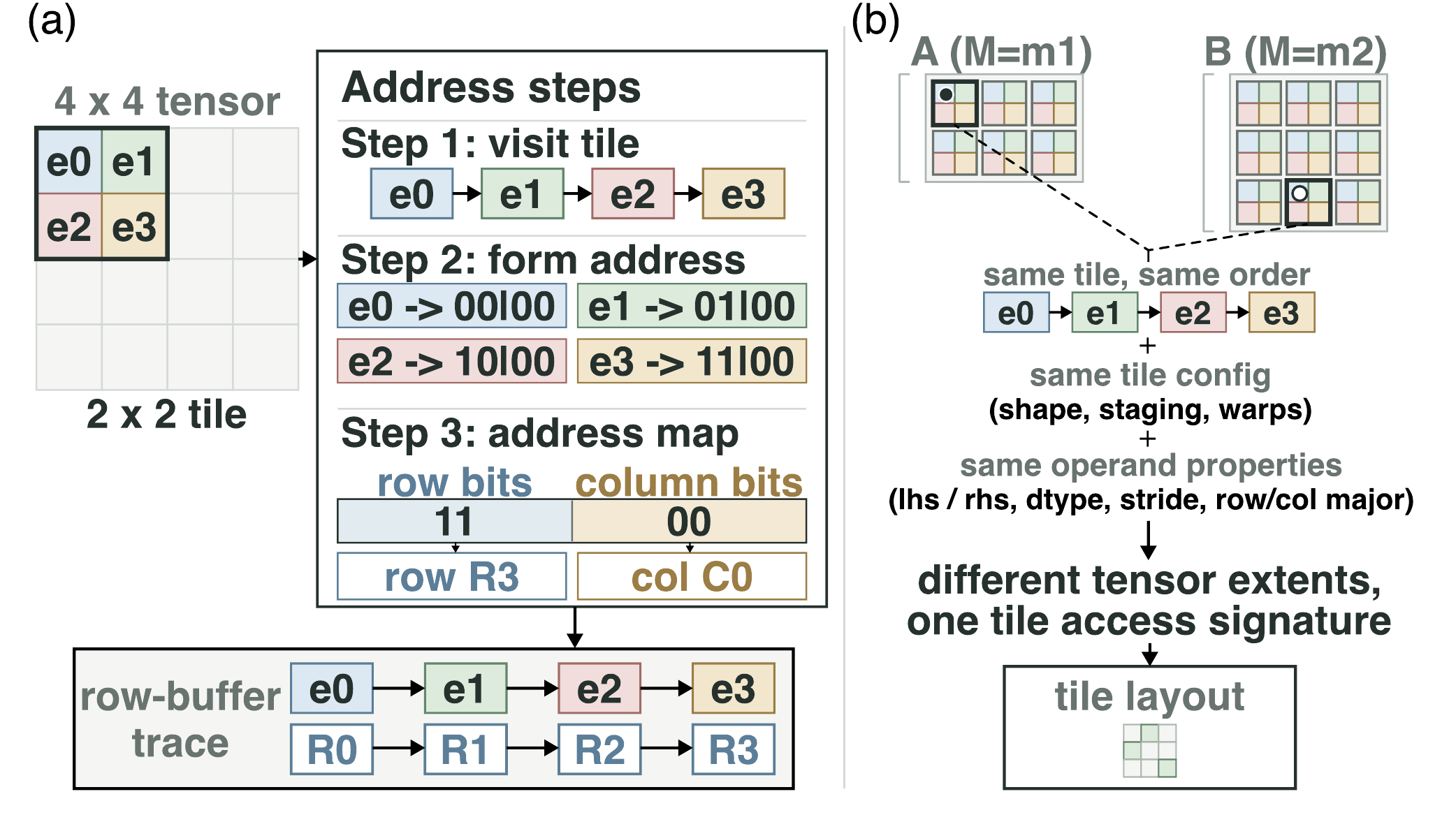}
\caption{The running $2\times2$ tile makes placement and reuse visible.
(a) The event strip above the tile is the complete ordered tile-local access sequence; the lower rail
shows the row-buffer states it drives under the conventional map.
(b) Instances at different origins share one access signature and use one
tile-to-DRAM layout.}
\Description{Panel (a) shows four ordered events in a two-by-two tile,
conventional placement under one fixed DRAM address mapping. The lower trace
separates the event order from the row-buffer state, showing
$\mathrm{closed}\rightarrow R0\rightarrow R1\rightarrow R2\rightarrow R3$;
the last three transitions conflict. Panel (b) shows
two tensors with different extents along M, each composed of the same tile.
Selected instances at different origins share the same access signature and are
shown as identical tile-local event views, revealing one shared tile-to-DRAM layout.}
\label{fig:bg-tile-conflict-reuse}
\end{figure}

The local stack nevertheless retains banked, row-buffered DRAM service, so
realizing the potential of this high-bandwidth path depends on the row sequence
induced by DRAM placement. Across our 108 full-trace MoE FFN cases, conventional placement yields a 75.7\%
request-weighted row-conflict rate (\secref{sec:eval-rq1-row-conflicts}).
\figref{fig:bg-tile-conflict-reuse}(a) isolates the mechanism at
tile granularity: a repeated tile-local difference maps to repeated row changes,
motivating explicit compiler control through a tile-to-DRAM layout.

\subsection{Stable Specialization Granularity in MoE FFNs}
\label{sec:bg-stable-specialization}

Application-specific synthesis constructs a workload-tuned controller mapping
from representative address differences \cite{rotaru2025xordrammapping}, shared
by all tensor and kernel requests reaching
the controller. Two properties of MoE FFN traffic, together with the cost of
exhaustive trace-wide controller-map synthesis, make the complete trace an
impractical specialization unit.

Heterogeneity appears within a single GEMM. The $A$ (lhs) operand advances over
row and reduction coordinates, the $B$ (rhs) operand over reduction and
output-feature coordinates, and both are loaded in the same reduction loop. One
global mapping must therefore serve two distinct address-difference
distributions \cite{tritonGroupedGemmTutorial,tritonMatmulTutorial}.

The complete trace also depends on runtime composition: routing sets expert-local
workloads, while grouped GEMM uses runtime $(M,N,K)$ values to form and schedule
tiles across problems \cite{hwang2023tutel,tritonGroupedGemmTutorial,
cutlassGroupedScheduler}. Expert-local $M$ changes the number of tile instances,
their source-interval bases, and the interleaving of requests from different
tiles \cite{cutlassGroupedScheduler,tritonGroupedGemmTutorial,tritonMatmulTutorial}.

Trace-wide synthesis is also expensive: because each outer mapping candidate
induces a separate trace-dependent fixed-domain problem, exhaustively searching
the evaluated mapping families against the Llama~4 full trace requires on the
order of $10^{14}$ evaluations of candidate copy events before any fixed-domain
solve, and every new phase, problem shape, or autotuned tile configuration
presents a fresh workload-wide target
(\appref{app:optimization-complexity}).

By contrast, the tile-local view is invariant. Once addresses are expressed as
source-interval offsets, tile configuration and operand properties fix one
ordered tile-local access sequence across instance origins and evaluated $M$
values (\figref{fig:bg-tile-conflict-reuse}). The tile access signature records
these fields together with the resulting sequence, so TileDRAM selects one
tile-to-DRAM layout per signature
(\secref{sec:tiledram-dram-domain-gf2-layouts}).

\section{Overview}
\label{sec:overview}

\begin{figure*}[!t]
\centering
\includegraphics[width=\textwidth]{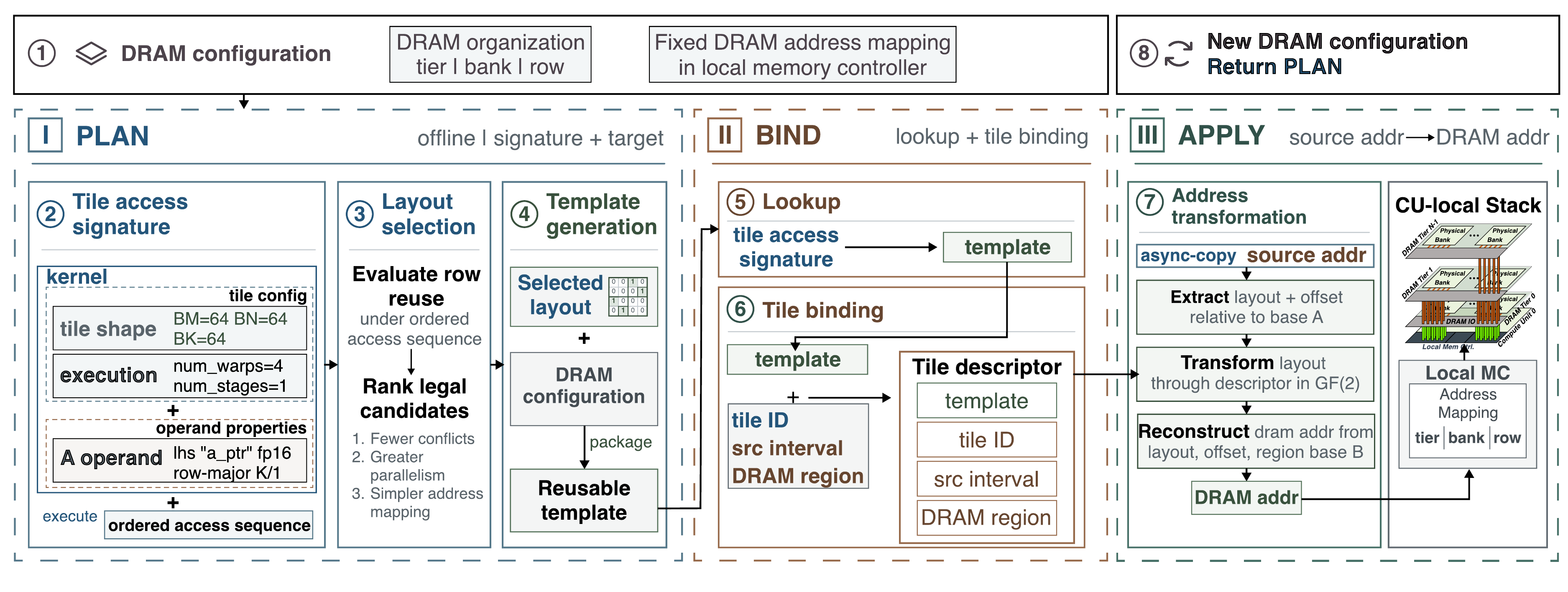}
\caption{Overview of TileDRAM's target-scoped PLAN--BIND--APPLY lifecycle. The
upper band fixes the DRAM configuration; PLAN combines it with a selected
layout to package a reusable template; \textsc{BIND} and \textsc{APPLY} then
perform tile binding and per-event address transformation under the fixed
mapping.}
\Description{The figure reads left to right across PLAN, BIND, and APPLY under
a target contract. PLAN combines the tile access signature (tile
configuration, operand properties, and the ordered tile-local access sequence) with the target, selects one
legal layout through ordered requests and row reuse, ranking candidates by row conflicts,
source-sequence domain coverage, and address-generation complexity, combines it
with the DRAM configuration, and packages the result as a reusable template.
BIND looks up the PLAN-produced template for each tile access signature and
combines it with the tile ID, source interval, and the DRAM region to form a tile
descriptor. APPLY
shows GF(2)-based address transformation from an asynchronous-copy source
address to a backend address. The transformed address is combined with the
region base to form the backend address. A new DRAM configuration returns
to PLAN to reselect the layout; BIND and APPLY retain the same interfaces.
}
\label{fig:overview}
\end{figure*}

TileDRAM realizes compiler-controlled DRAM placement as a target-scoped
PLAN--BIND--APPLY lifecycle (\figref{fig:overview}).
\stepnum{1}~The DRAM configuration fixes the DRAM organization and the
controller-owned DRAM address mapping.
\stepnum{2}~A tile access signature combines tile configuration, operand
properties,
and the ordered tile-local access sequence.
\stepnum{3}~Layout selection evaluates layout candidates over the ordered requests for row reuse and
ranks legal candidates by fewer row conflicts, greater parallelism, and lower
address-generation complexity.
\stepnum{4}~The selected layout is combined with the DRAM configuration
and packaged as a reusable tile-to-DRAM Linear Layout template.
\stepnum{5}~\textsc{BIND} looks up the \textsc{PLAN}-produced
template for each tile access signature.
\stepnum{6}~\textsc{BIND} then combines the referenced
template with the tile ID, source interval,
and the DRAM region to construct a tile descriptor.
\stepnum{7}~During execution, \textsc{APPLY} transforms each source
address using a GF(2)-based address transform, then uses the descriptor's
region base to form the backend address.
\stepnum{8}~A new DRAM configuration reruns \textsc{PLAN};
\textsc{BIND} and \textsc{APPLY} remain unchanged.

\section{Tile-to-DRAM Linear Layouts}
\label{sec:dram-domain-linear-layouts}

This section develops tile-to-DRAM Linear Layouts as an \fTwo-linear
representation of compiler-controlled DRAM placement. For each tile access
signature, we formalize how a
candidate layout and the signature's ordered tile-local access sequence jointly determine
row-buffer behavior. TileDRAM formulates layout selection as a row-locality
optimization subject to the target DRAM address mapping and placement constraints.
Each admitted signature yields a reusable tile-to-DRAM layout template for the target
contract, used to bind logical tiles to concrete DRAM regions.

\subsection{Tile-to-DRAM Linear Layout Object}
\label{sec:tiledram-dram-domain-gf2-layouts}

A tile-to-DRAM Linear Layout maps each layout unit to a row-buffer domain, a
row within that domain, and a within-row position. We call the hardware unit
that independently holds one open row a \emph{row-buffer domain}, or simply a
domain; in the evaluated 3D organization, a domain is one (tier, bank) pair,
but the construction needs only whatever unit the target tracks independently.

TileDRAM reuses the labeled \fTwo algebra of GPU Linear Layouts for a separate
logical-tile-to-DRAM placement map \cite{linearlayouts2025}. Triton distributed
layouts map hardware-resource coordinates to logical tensor coordinates;
TileDRAM maps normalized tile coordinates to DRAM placement coordinates.
Each recorded access carries a source address $s_i$ in the global space.
Let $A$ denote the source-interval base and $g$ the granularity of one access.
Normalization extracts the layout-unit index
$\ell_i=\lfloor(s_i-A)/g\rfloor$ and within-unit byte offset
$\delta_i=(s_i-A)\bmod g$. Writing $r_i\in\fTwo^n$ for the \fTwo encoding of
$\ell_i$,
a reusable \emph{tile-to-DRAM Linear Layout} is the strictly linear map
\begin{equation}
  \begin{bmatrix}p_i\\d_i\\z_i\end{bmatrix}
  =W_{\mathrm{sem}}r_i
  =\begin{bmatrix}W_C\\W_D\\W_R\end{bmatrix}r_i,
  \qquad W_{\mathrm{sem}}\in\fTwo^{m\times n}.
  \label{eq:semantic-dram-linear-layout}
\end{equation}
The $W_D$ slice determines the row-buffer domain, $W_R$ the row within that
domain, and $W_C$ the within-row position of the layout unit.
Equal $W_D$ images therefore share a domain, while equal $(W_D,W_R)$ images
target the same row. The domain's current open-row state determines whether a
request is a hit, a conflict, or an opening miss. This expresses row locality in
the same labeled-map algebra as compute-side layouts.

TileDRAM selects one layout per \emph{tile access signature}, written
$\theta=(\kappa,\rho,\chi)$. Here, $\kappa$ denotes the tile configuration,
comprising tile shape $(B_M,B_N,B_K)$, pipeline-stage count
(\texttt{num\_stages}), and warp count (\texttt{num\_warps}); $\rho$ denotes
the operand-property tuple, comprising operand position (\texttt{lhs} or
\texttt{rhs}), data type (\texttt{dtype}), source-stride pattern
(\texttt{stride}), and source layout (row-major or
column-major); and $\chi$ denotes the ordered tile-local access sequence. A
representative
execution yields the normalized source sequence $S_\theta$ for this signature,
which is used for layout selection.

\begin{figure}[!t]
\centering
\includegraphics[width=\columnwidth]{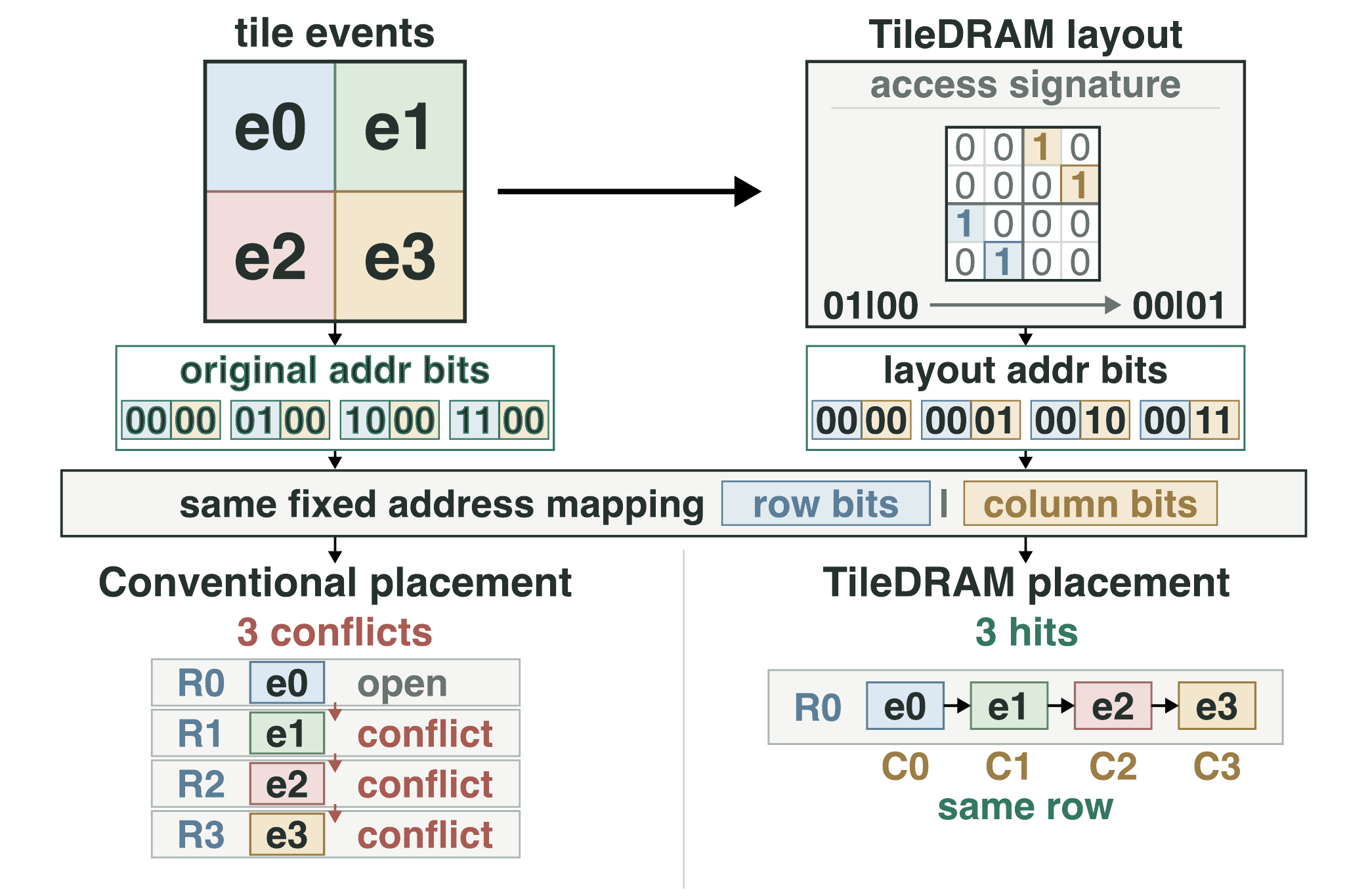}
\caption{In a single-row-buffer projection of the
\figref{fig:bg-tile-conflict-reuse} stream, a tile-to-DRAM Linear Layout places
$e_0$--$e_3$ in row $R_0$ at columns $C_0$--$C_3$ under the same fixed address
mapping, turning the conventional three conflicts into three row hits.}
\Description{The same ordered two-by-two tile and TileDRAM layout feed two
address-field streams through one fixed address mapping. The compact card uses the
tile access signature as the lookup key and shows its selected template; the
representative field transform maps $01|00 \mapsto 00|01$. In the original stream, the
left row-bit field changes as 00, 01, 10, 11 while the right column-bit field
stays 00; in the layout stream, the left field stays 00 while the right field
changes as 00, 01, 10, 11. The lower comparison uses one row-buffer
organization: conventional placement visits rows R0, R1, R2, and R3, while
TileDRAM placement keeps all four events in row R0 at C0, C1, C2, and C3.}
\label{fig:layout-row-semantics}
\end{figure}

Each live layout-unit index $\ell_i$ is injectively encoded as its
coordinate $r_i$; the resulting
sequence $S_\theta=(r_0,\ldots,r_{N-1})$ is the ordered input to layout
selection. We evaluate this construction on tiled GEMM under the conditions stated in
\appref{app:generality-conditions}.
\figref{fig:layout-row-semantics} shows the single-row-buffer projection of the
\figref{fig:bg-tile-conflict-reuse} stream under a tile-to-DRAM Linear Layout.

The formal treatment below retains the target's independent row-buffer domains
and their separate open-row states. The map in
Eq.~\eqref{eq:semantic-dram-linear-layout} defines a static DRAM placement, whereas
row-buffer outcomes depend on request order. TileDRAM therefore augments the
layout with ordered row-buffer state semantics.

\subsection{Ordered Row-Buffer Semantics and Row-Reuse Kernel Reduction}
\label{sec:tiledram-projection-kernel-optimization}

Immediately before request $i$, each domain either holds one open row or is
closed; write $\sigma_i(d)$ for that state, with $\bot$ denoting closed, and
let every domain start closed. Assuming the controller retains open rows across
requests, the row-buffer transition over $S_\theta$ is
\begin{equation}
  \epsilon_i=\begin{cases}
    \mathrm{miss},&\sigma_i(d_i)=\bot,\\
    \mathrm{hit},&\sigma_i(d_i)=z_i,\\
    \mathrm{conflict},&\text{otherwise},
  \end{cases}
  \qquad \sigma_{i+1}=\sigma_i[d_i\mapsto z_i].
  \label{eq:main-row-state-transition}
\end{equation}
An opening miss is the ``row closed'' case of classic memory schedulers
\cite{mutlu2007stfm} and pays one activation; a conflict requires a precharge and an activation before that request can be served; a hit requires neither. TileDRAM's
source-order objective minimizes conflicts first; the remaining selection
criteria are defined in \secref{sec:dram-layout-selection}.

For a fixed domain projection $W_D$, the \emph{row-reuse kernel}
\[
  K=\ker\!\begin{bmatrix}W_D\\W_R\end{bmatrix}
\]
collects the source-coordinate differences that change neither the domain nor
the row label. For request $i$ with same-domain predecessor $j$, the request is
an intrinsic source-order hit iff $r_i\oplus r_j\in K$. Requests with no
same-domain predecessor are opening misses. Intervening requests to other domains
do not change this state. Thus, the source-order hit/miss/conflict classification is fixed by the source
order and layout before scheduler effects. For the two-domain
realization of this stream in \appref{app:xor-weighted-diff}
(Table~\ref{tab:redline-kernel-example}), the two same-domain predecessor
differences are both $001$, so the
kernel $K=\mathrm{span}\{001\}$ contains both and yields the two hits used in
the formal count. Let
$w_{W_D}(\Delta)$ be the
number of times difference $\Delta$ occurs between consecutive same-domain
requests. Counting hits therefore reduces to summing $w_{W_D}$ over $K$.
The target organization bounds how many bits may name a domain ($b$) and how
many may name a row within it ($r_{\mathrm{row}}$). When the joint domain--row
map $[W_D;W_R]$ has
rank $b+r_{\mathrm{row}}$, choosing those widths fixes
$c=n-b-r_{\mathrm{row}}$, the dimension of the row-reuse kernel.

This reduction requires three properties, each formalized and proved in the
appendix. \emph{Sequence recurrence}: a signature determines one ordered
coordinate sequence, so one representative profile applies across its admitted
instances (\appref{app:generality-conditions}). \emph{Mapping consistency}:
within the admitted bit-selection-and-XOR family, DRAM placement and the target
DRAM address mapping compose into an affine \fTwo\ map whose domain- and
row-equality relations agree with those resolved by the controller
(\appref{app:dram-domain-semantics}). \emph{Binding legality}: reused
bindings retain the signature's source interpretation and request formation,
realize those labels consistently, and satisfy the placement bounds, so every
admitted tile placement reproduces the selected row behavior
(\appref{app:materialization-legality}).

For a fixed $W_D$, row sublayouts whose joint domain--row maps have the same
kernel induce the same domain-local row-equality relation, so the inner search
can operate on subspaces $K$ rather than on matrix representations of $W_R$.
Once the three properties hold, maximizing row hits over placements therefore
reduces exactly to choosing the maximum-weight $c$-dimensional subspace, giving the fixed-domain objective
\begin{equation}
  \max_{K\subseteq\ker(W_D),\ \dim K=c}
  \sum_{\Delta\in K}w_{W_D}(\Delta).
  \label{eq:layout-fixed-domain-kernel}
\end{equation}
Rotaru et al.~formulate the same subspace-selection problem
\cite{rotaru2025xordrammapping}.

The two formulations reverse which map is the decision variable: Rotaru
et al. synthesize the controller address map with data placement fixed, whereas
TileDRAM synthesizes data placement with the controller mapping fixed.  This is
a reversal of the control variable, rather than an algebraic duality. GPU Linear Layouts make that decision variable
expressible: the layout algebra that compilers already use for lanes and registers
now extends to row-buffer domains and rows, turning DRAM placement into a
compiler-visible decision variable. Taking differences over tile-local access
coordinates rather than a flat trace then lets one solution serve every tile
sharing the signature.

The objective optimizes the layout under each signature's ordered tile-local access sequence. The predecessor, hit, and coordinate-lifting
derivations appear in
Appendices~\ref{app:xor-weighted-diff}--\ref{app:fixed-domain-solver}.

\subsection{Layout Synthesis and Selection}
\label{sec:dram-layout-selection}

Because $W_D$ determines predecessors within each row-buffer domain and thus
$w_{W_D}$, Eq.~\eqref{eq:layout-fixed-domain-kernel} is evaluated for each fixed
domain projection. The resulting five-stage construction forms the
layout-selection
core of \textsc{PLAN} and produces an executable layout from $S_\theta$ and
the target organization.

\begin{enumerate}[leftmargin=*,topsep=0.25em,itemsep=0.15em,parsep=0pt]
\item \textbf{Row-buffer-domain projection.} Enumerate a finite target-specific
family of row-buffer-domain sublayouts $W_D$. Each candidate assigns source coordinates to the
row buffers independently tracked by the organization; the catalog
holds bit-selection and constrained structured-XOR candidates.

\item \textbf{Kernel construction.} Form predecessor differences within each
row-buffer domain and solve Eq.~\eqref{eq:layout-fixed-domain-kernel} for a
$c$-dimensional row-reuse kernel $K$.

\item \textbf{Layout completion.} Reconstruct a row sublayout $W_R$ whose joint
row-buffer-domain and row map has kernel exactly $K$ using the sparse construction of
Rotaru et al.~\cite{rotaru2025xordrammapping}, then add $W_C$ to bring
$W_{\mathrm{sem}}$ to full column rank.

\item \textbf{Physical realization.} Express $W_{\mathrm{sem}}$ in the target's
address basis to obtain the executable $W_{\mathrm{phys}}$, then derive its
byte-placement rule. In this basis, the fixed address mapping recovers the
intended row-buffer-domain and row labels, and ordinary bit operations generate the backend
offset. Retained candidates are injective on live coordinates and satisfy
mapping consistency (\appref{app:dram-domain-semantics}) together with alignment,
capacity, CU-local aperture bounds, and allocation-policy constraints
(\appref{app:materialization-legality}).

\item \textbf{Ordered selection.} Run the row-state transition of
Eq.~\eqref{eq:main-row-state-transition} on every retained layout. Selection
first minimizes source-order conflicts. Among equal-conflict layouts, it favors
greater source-sequence domain coverage, measured by
$\mathsf{cov}(S,W_D)$ in Appendix~\ref{app:domain-parallelism}, and then lower
nonzero-term complexity of the executable address map.
\end{enumerate}

The outer loop exhaustively enumerates the finite domain-projection family. For
each $W_D$, the exact route finds the optimum of
Eq.~\eqref{eq:layout-fixed-domain-kernel}, while the greedy route returns a feasible kernel
(\appref{app:fixed-domain-solver}). Reconstruction and legality filtering
retain only candidates satisfying live-coordinate injectivity, target mapping
consistency, and binding legality. The retained candidates are ranked
exhaustively by the lexicographic objective
(Appendices~\ref{app:xor-reconstruction}--\ref{app:xor-equivalence})
within the enumerated candidate family. The selected $W_{\mathrm{sem}}$ is realized in the target
address basis and reused across legal tile placements.

\section{Compiler and Runtime Realization}
\label{sec:compiler-runtime-realization}

\begin{figure}[!b]
\centering
\includegraphics[width=\columnwidth]{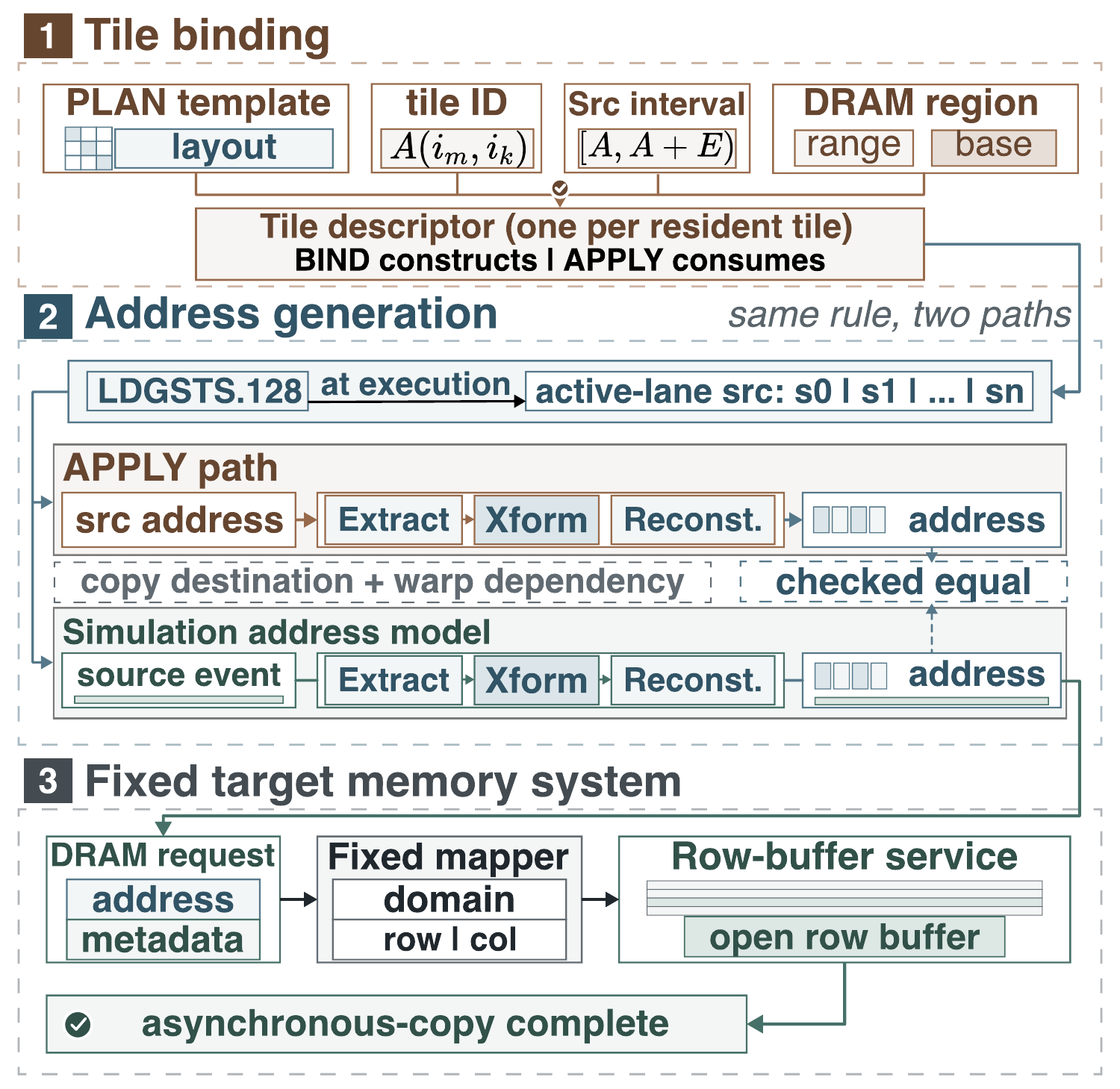}
\caption{Compiler/replay address equivalence. Compiler-side \textsc{BIND} looks up the
\textsc{PLAN}-produced layout template; runtime \textsc{BIND} combines it with the
tile ID, source interval, and DRAM region to form a tile descriptor. The
compiler-generated \textsc{APPLY} path and replay apply the same address transformation to
active-lane source addresses, with replay using the corresponding addresses carried
by asynchronous-copy events, before fixed address mapping and row-buffer service.}
\Description{The figure has three stages. Compiler-side \textsc{BIND} performs template lookup for a
\textsc{PLAN} layout template and emits binding metadata; runtime \textsc{BIND}
obtains that template and combines it with the concrete tile ID, source interval,
and DRAM region into a tile descriptor for that region. Address
transformation shows an APPLY path and a
replay path using the same address transformation; the APPLY path is
generated once from the template, while the runtime tile descriptor is an
execution input to APPLY and replay. Their backend addresses are checked
for equality while the copy destination and dependency remain unchanged. A DRAM
request then passes through the fixed address mapper and DRAM row-buffer service
before completing the copy dependency.}
\label{fig:tiledram-unified-realization}
\end{figure}

The target-specific tile-to-DRAM layout template developed in
\secref{sec:dram-domain-linear-layouts} is realized through the three-stage
PLAN--BIND--APPLY lifecycle. \textsc{PLAN} selects and realizes the template.
\textsc{BIND} spans compilation and runtime: compiler-side
\textsc{BIND} performs template lookup, emits binding metadata, and marks the
corresponding accesses for \textsc{APPLY}; runtime \textsc{BIND} uses that
metadata to construct a tile descriptor for a resident tile (a concrete tile
instance placed in a DRAM region). During execution, \textsc{APPLY} consumes that descriptor to
transform each active lane's source address; GPU execution and replay use the
same address transformation.
In the evaluated Ampere-class
realization, operand-tile loads use cp.async lowered to LDGSTS.128.
\figref{fig:tiledram-unified-realization} isolates this compiler/replay address
equivalence, and \appref{app:compiler-interface} shows the programmer-facing
kernel entry.

\subsection{Semantic Extraction and Offline Template Construction}
\label{sec:tiledram-compiler-realization}

\paragraph{Programming interface.}
TileDRAM does not require programmers to name DRAM banks, rows, or
address-map bits. The programmer-facing annotation supplies problem, tile,
execution, and operand properties, while the kernel body remains ordinary
Triton/Gluon code (\appref{app:compiler-interface}); DRAM placement remains
TileDRAM-managed rather than part of the kernel's logical tile specification.

\paragraph{Semantic extraction.}
From the tiled-kernel representation, the compiler derives the tile
configuration and operand properties. A recorded representative active-lane
sequence supplies the ordered tile-local access sequence; together, these
fields form the tile access signature. Normalization relative to the source
interval (\secref{sec:dram-domain-linear-layouts}) yields the representative
normalized source sequence used by \textsc{PLAN}. For each admitted signature and
target contract, \textsc{PLAN} synthesizes the layout and its target-specific layout
realization, publishing one reusable template keyed by the signature.

\subsection{Template Binding and Address Lowering}
\label{sec:tiledram-template-binding}

\paragraph{Tile binding (\textsc{BIND})}
On the compiler side, \textsc{BIND} performs template lookup for each admitted
tile access signature and records a reference to the corresponding
target-specific layout template produced by \textsc{PLAN}, together with the
rule for deriving the tile ID from the execution context. A tile ID identifies
an operand tile within the current GEMM instance. For GEMM, the lhs and rhs tile
IDs are represented by the operand-tile coordinates $(i_m,i_k)$ and
$(i_k,i_n)$, respectively. When the tile is placed in a DRAM region, the compiler-defined tile-ID rule yields
the concrete tile ID from the current execution context, and runtime
\textsc{BIND} obtains the referenced template from this binding metadata and
combines it with the tile ID, source interval $[A,A+E)$, and concrete
DRAM region to construct a tile descriptor for the resident tile. Each
admitted resident tile uses one tile access signature and one corresponding
target-specific layout template throughout its residency, so its descriptor
remains fixed and is reused by accesses to the same resident tile.
Thus, lhs accesses from CTAs
that differ only in $i_n$, and rhs accesses from CTAs that differ only in $i_m$,
resolve to the same tile descriptor.
At execution, runtime \textsc{BIND} provides the descriptor for the
current resident tile as input to \textsc{APPLY}.

\paragraph{Address lowering (\textsc{APPLY})}
At compilation, the compiler places \textsc{APPLY} after the source pointer
tensor is formed by address arithmetic for each admitted tile access. At execution, the path
uses the tile descriptor for the current resident tile to
transform each active lane's source address for its
16-byte LDGSTS.128 transfer. The transformation follows three ordered stages:
\emph{Extract} each source address relative to the source interval, yielding the
layout-unit index and within-unit byte offset; \emph{Transform} the layout
unit through the descriptor's target-specific layout realization; and
\emph{Reconstruct} the backend address from the transformed unit,
unchanged byte offset, and region base before the copy. Per-lane integer
and pointer operations realize this software-selected placement through ordinary
address generation, requiring no new address-remapping hardware or ISA support.
The formal address contract is given in \appref{app:dram-domain-semantics}, and
the binding metadata and address-lowering sequence appear in
\appref{app:compiler-interface}. Only operand accesses with a tile descriptor
are redirected to their assigned DRAM regions, while ordinary
loads and stores retain the conventional memory path.

\subsection{Runtime Replay and Simulator Execution}
\label{sec:tiledram-runtime-simulator-realization}

Simulator replay represents each dynamic tile access as an ordered tile-local access sequence of
active-lane asynchronous-copy events (LDGSTS.128) carrying source address $s_i$ and event
metadata $\eta_i$. Here, $\eta_i$ denotes the event's identity, destination,
dependency, and request attributes. Runtime scheduling may interleave ordered tile-local access sequences
from resident tiles while preserving the event order within each sequence.
Each sequence carries the descriptor for its resident tile as
input to the \textsc{APPLY} path.

\label{sec:design-runtime-projection}
\label{sec:cross-path-address-equivalence}
\label{sec:design-simulator-observation}
For each asynchronous-copy event, the simulator evaluates the same address
transformation specified by \textsc{APPLY} using the descriptor for the
sequence's resident tile. It carries the resulting backend address
together with the event metadata into replay. Replay uses the descriptor's
template and placement while preserving event metadata and dependency. The
production compiler lowering and simulator-side transformer implement one
common address function, checked against an independent reference evaluator.
\appref{app:artifact-configuration} reports the covered configurations,
address-function families, and byte addresses. Each transformed
asynchronous-copy event forms one CU-local DRAM request in the Ramulator2
replay. Ramulator2's fixed address mapper and controller advance row-buffer
state and return service latency to GPGPU-Sim, which applies it to the
originating asynchronous-copy dependency.

\section{Evaluation}
\label{sec:evaluation}

\subsection{Experimental Setup}
\label{sec:eval-questions-setup}

\paragraph{Simulation and replay protocol.}
We evaluate TileDRAM with traces collected from Triton/Gluon tiled GEMM kernels using NVBit
\cite{triton2019,tritonGluon,villa2019nvbit} and replayed through Accel-Sim/Ramulator2
\cite{khairy2020accelsim,luo2023ramulator2} on one CU and its paired 3D DRAM stack.
Each
baseline--TileDRAM pair holds the logical event trace, target organization,
the initial closed-row state, and refresh/precharge policy fixed while varying
DRAM placement. Layout selection uses the ordered tile-local access sequence, whereas
replay counters follow FR-FCFS service order. Before replay, we verify that the
compiler lowering and simulator transformer produce the same addresses as the
canonical byte-address function in \appref{app:dram-domain-semantics}.

The evaluation covers 108 full-trace MoE FFN cases spanning Qwen3
\cite{qwen3report2025}, DeepSeek-V3 \cite{deepseekv3report2024}, Qwen3.5-35B-A3B
\cite{qwen3535ba3b2026}, Kimi K3 \cite{kimik3report2026}, Hunyuan-A13B
\cite{hunyuana13b2025}, and Llama 4 \cite{llama4herd2025}, across up/gate and
down projections and eight tile shapes. The same 108
kernel configurations provide hardware-cost calibration; a separate sweep
covers 36 full-trace stream--organization pairs.
All reported TileDRAM layouts use the same greedy selection route
(\appref{app:fixed-domain-solver}). We report row-conflict rates from pooled
row events and speedup as the geometric mean of per-case
baseline-to-TileDRAM cycle ratios. Per-case synthesis time and peak resident
memory measure offline layout-selection cost; address-transformation cost is
measured separately on RTX~3090 hardware.
The primary 3D DRAM configuration uses 8 tiers, 16 banks, 13 row bits, and
3 column bits with RoBaTiCoCh address mapping, OpenRowPolicy, and FR-FCFS
scheduling.

\subsection{RQ1: Row-Buffer Cost of Conventional DRAM Placement}
\label{sec:eval-rq1-row-conflicts}

\begin{figure}[!t]
\centering
\includegraphics[width=0.85\columnwidth]{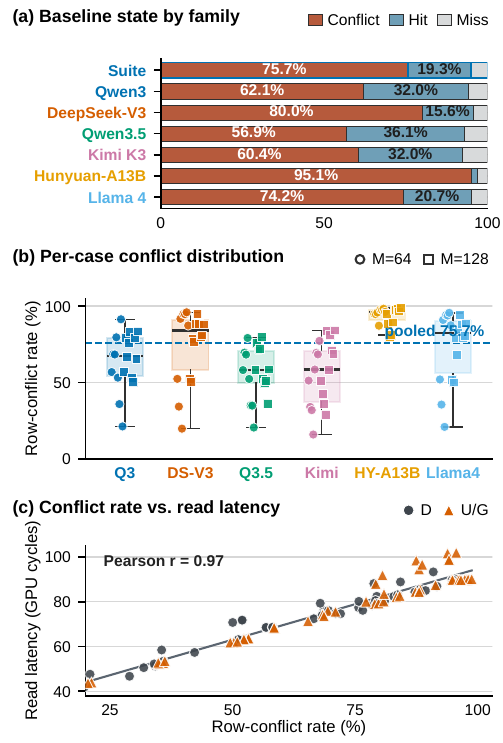}
\caption{Conventional placement across 108 full-trace MoE FFN cases.
(a) Request-weighted row states by family; (b) per-case conflict distributions;
and (c) conflict rate versus average read latency.}
\Description{The single-column figure has three vertically arranged panels.
Panel (a) shows request-weighted conflict, hit, and miss shares for the suite
and six model families. Panel (b) shows six family-wise boxplots and all 108
case conflict rates, using circles for M equals 64 and squares for M equals
128; a dashed line marks the pooled 75.7 percent conflict rate. Panel (c) plots
each case's conflict rate against average read latency, using circles for down
projections and triangles for up and gate projections; the fitted trend has a
Pearson correlation of 0.97.}
\label{fig:baseline-pathology}
\end{figure}

\begin{figure*}[!t]
\centering
\includegraphics[width=\textwidth]{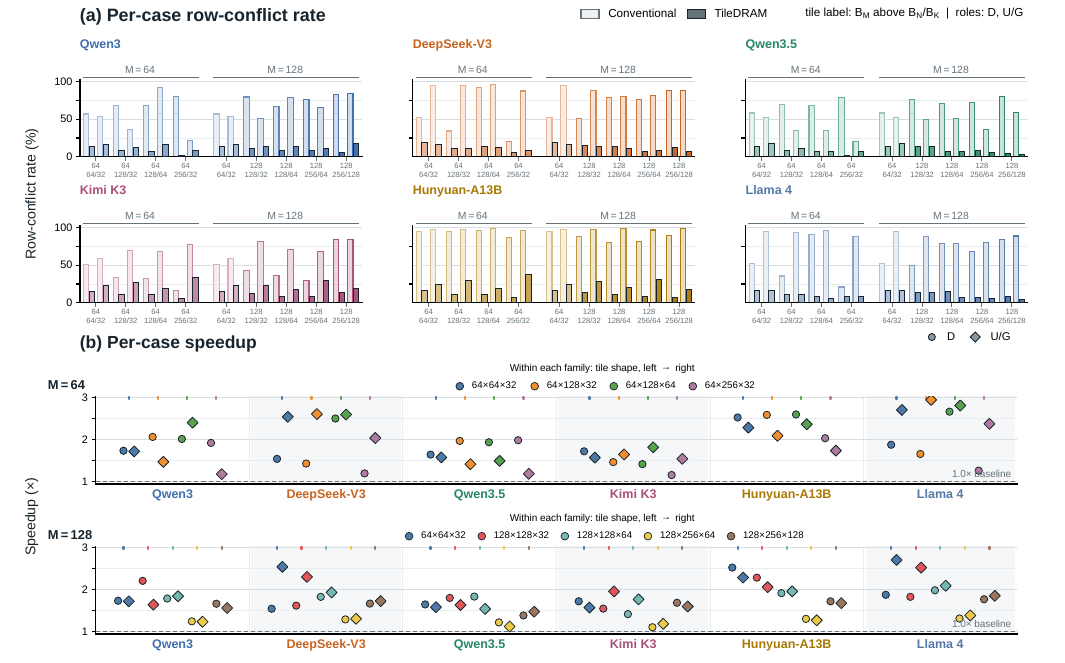}
\caption{Row-conflict rate and speedup for 108 MoE FFN cases.
(a) Paired conflict-rate bars per case; the headers identify $M$ and the
labels below each group show $B_M\!\times\!B_N\!\times\!B_K$. (b) Speedup grouped by
family, faceted by $M$; within each family, color and x-position identify tile
shape.}
\Description{Six small bar charts show paired Conventional and TileDRAM
row-conflict rates for eighteen cases per model family; $M$ headers and the
short labels below each group identify $B_M\!\times\!B_N\!\times\!B_K$. Two
dot plots below show speedup for M=64 and M=128, with model family on the
x-axis, tile shape encoded by labeled colors and fixed positions, and
projection type by marker shape.}
\label{fig:completed-results}
\end{figure*}

\figref{fig:baseline-pathology} reports a 75.7\% request-weighted row-conflict
rate across all 108 cases. Per-case rates span 15.9--98.8\% and family-pooled
rates 56.9--95.1\%, while the $M=64$ and $M=128$ groups differ little
(73.9\% and 76.7\%).
The remaining variation follows model family, projection type, and tile
configuration, whose ordered tile-local access sequences induce different domain
and row trajectories under the target DRAM address mapping. The baseline pathology
reflects two interacting properties of GEMM execution. First, its operand-local
address structure is misaligned with the controller's fixed address-field
interpretation: regular tile-coordinate changes repeatedly revisit one
row-buffer domain at different rows. Second, \texttt{cp.async} exposes these loads through a
lane- and warp-interleaved issue order rather than a complete tensor-row scan,
separating accesses that could reuse an open row. The mapping fixes the
domain/row relation, while this issue order determines when it is exposed,
producing repeated precharge--activate transitions and row conflicts.
FR-FCFS can reorder ready requests, but cannot change the assigned domain or row.

Average read latency correspondingly ranges from 41.1 to 101.9 GPU cycles and
closely tracks row-conflict rate (Pearson $r=0.97$). Because placement
shapes the row sequence each domain sees, the suite-wide prevalence
of conflicts motivates treating DRAM placement as a software optimization
target, while the case-level variation motivates layout selection per
tile access signature.
\subsection{RQ2: Row-Buffer Locality and Performance}
\label{sec:eval-row-layout}

TileDRAM converts a large fraction of suite-wide row conflicts into row hits. Across the
108 cases in \figref{fig:completed-results}(a), the request-weighted
row-conflict rate falls from 75.7\% to 15.2\%, an 80.0\% pooled reduction, while
the row-hit rate rises from 19.3\% to 81.6\%. Conflict rates fall for every
model family, $M$ value, and tile shape. Family-pooled TileDRAM conflict rates remain within
10.8\%--21.2\%, compared with 56.9\%--95.1\% under conventional placement.
Layout selection per tile access signature therefore adapts to
the access-pattern heterogeneity measured in RQ1 and produces consistently high
row locality.

TileDRAM reshapes each operand stream's row-buffer-domain and row sequence.
Many same-domain requests therefore hit open rows, replacing a substantial
fraction of precharge--activate transitions with row-hit service. Across cases, the median reductions in ACT and PRE command counts are
78.7\% and 82.2\%, respectively (IQRs: 65.7--84.6\% and 70.5--87.8\%). Request-weighted read latency falls from 81.0
to 40.8 GPU cycles, and foreground queue wait falls from 37.3 to 24.9 cycles.

Earlier DRAM completions release asynchronous-copy dependencies sooner, reducing
pooled shader memory-stall cycles from 2.086 to 0.962 billion. Speedup
then spans 1.10$\times$--2.95$\times$, with a 1.77$\times$ geomean across all
108 cases. Across cases, the percentage-point reduction in row-conflict
rate tracks the cycle reduction in read latency ($r=0.96$); relative latency
and stall reductions track each other ($r=0.89$), and relative stall reduction
tracks speedup ($r=0.90$).

The performance range reflects the fraction of each baseline execution occupied
by memory stalls. For the $64{\times}128{\times}64$ tile, the row-conflict rate
falls from 87.5\% to 11.9\%; memory stalls occupy 92.0\% of baseline cycles, and
cases using this tile achieve a 2.17$\times$ geomean speedup. For
$128{\times}256{\times}64$, conflicts fall from 74.7\% to 12.4\%; with a 58.6\%
baseline stall share, cases using this tile achieve a 1.25$\times$
geomean speedup. Thus, TileDRAM consistently shortens DRAM service, while
the baseline memory-stall share sets the headroom for speedup.

\subsection{RQ3: Retargeting Across 3D DRAM Organizations}
\label{sec:eval-rq3-scheme-sensitivity}

TileDRAM preserves row locality when retargeted across the evaluated 3D DRAM
organizations. We evaluate six full-trace FFN streams with the
$64{\times}64{\times}32$ tile across six target organizations, for 36 paired
replays. TileDRAM reselects a legal tile-to-DRAM layout for each target
organization.

\begin{figure}[!t]
\centering
\includegraphics[width=\columnwidth,clip,trim=0 0 0 0]{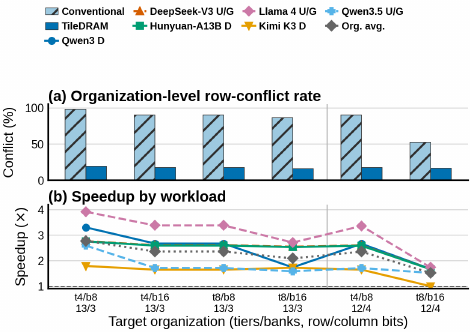}
\caption{Retargeting six full-trace FFN streams across six 3D DRAM
organizations. (a) Request-weighted row-conflict rate under conventional
placement and TileDRAM. (b) Per-stream speedup; gray diamonds report the
geometric mean across the six streams for each organization. D and U/G denote
down and up/gate projections.}
\Description{The figure compares row-conflict rate and speedup across
six 3D DRAM organizations. The upper plot uses paired conventional-placement
and TileDRAM bars. The lower plot shows six per-stream speedup lines and gray
diamonds for the geometric mean across streams within each organization.}
\label{fig:rq3-geometry}
\end{figure}

Across the sweep, the pooled row-conflict rate falls from 84.3\% to 17.6\%, a
79.2\% reduction (\figref{fig:rq3-geometry}). These reductions yield a
2.20$\times$ geomean speedup across the 36 paired replays.

Across organizations, six-stream geomean speedups range from
1.52$\times$ to 2.77$\times$, while pooled-cycle speedups range from
1.60$\times$ to 3.32$\times$. The resulting locality
is stable across the evaluated organizations, while baseline conflict headroom
and organization-specific DRAM service together determine how that locality
translates into cycles. Organization-specific layout selection therefore
preserves row locality even as the DRAM address mapping repartitions the
physical address space.

\subsection{RQ4: Cost of Layout Synthesis and Address Transformation}
\label{sec:eval-rq4-projection-cost}

RQ4 measures both the one-time cost of synthesizing a layout and the dynamic
cost of realizing it through compiler-lowered address transformations. Offline
synthesis runs once per case and produces the layouts for its tile access
signatures; the selected templates are reused across matching dynamic tile
instances. We record per-case solver wall
time and peak resident memory across 108 cases. Each case compares identity
address generation ($H1$) with the selected transformation ($H2$). The selected layouts lower to sparse XOR, grouped bit-permutation, and
identity paths using shifts, masks, XORs, and address arithmetic. We report the
overhead as
$\operatorname{median}(H2)/\operatorname{median}(H1)-1$ and aggregate the
corresponding ratios geometrically
(\appref{app:artifact-configuration}).

Offline synthesis has a median per-case runtime of 349.0~s and a median peak
resident memory of 227.0~MB. Across all 108 hardware-measured kernel configurations, the
selected transformations add 0.17\% geomean overhead relative to identity address
generation; the median is below 0.01\%, the 95th percentile is 0.57\%, and the
maximum is 4.93\%. Overall, 104 of 108 cases lie within $\pm$1\% and 106 lie
within $\pm$2\%; the Qwen3.5 down-projection configurations account for the
high-overhead tail.

\section{Related Work}
\label{sec:related}

\paragraph{3D DRAM Architectures and Accelerators.}
Vertical integration and hybrid bonding place high-bandwidth memory close to
compute, motivating architectures that exploit local memory--logic bandwidth
for LLM and MoE inference. H$^2$-LLM co-designs heterogeneous hardware and
dataflow around hybrid-bonded memory \cite{li2025h2llm}, HybridSpec combines
such memory with heterogeneous execution and speculative decoding
\cite{huang2026hybridspec}, Stratum uses tiered monolithic 3D DRAM and expert
placement for MoE serving \cite{pan2025stratum}, and Raptor demonstrates an
early-silicon 3D-DRAM accelerator \cite{nair2026raptor}. At industrial scale,
OpenAI's Jalape\~{n}o pairs compute core slices with local HBM4 slices in a
memory-sliced organization \cite{ho2026jalapeno}.
Other designs explore hybrid NVM--DRAM, processing-in-memory, and near-memory
parallelism \cite{he2026shyla,huang2025hdmoe,li2026hybridpim}. These systems
establish the architectural substrate for TileDRAM; TileDRAM instead exposes
how operand tiles are placed within local DRAM as a compiler-visible layout
decision.

\paragraph{Compute-Side Tile Layouts and Algebraic Layout Abstractions.}
Modern GPU programming systems make tiled tensor organization and layout
first-class abstractions: CUTLASS and Triton support tiled computation, while
CuTe provides composable tensor layouts and layout algebra
\cite{cecka2026cute,cutlassGemmApi,triton2019}. Graphene represents
multidimensional data and threads as hierarchically tiled first-class tensors,
expressing optimized GPU computations as mappings between data and thread tiles
\cite{hagedorn2023graphene}. Gluon exposes explicit tensor layouts for
low-level control, and Jalape\~{n}o uses Gluon to program its spatial
architecture \cite{ho2026jalapeno,tritonGluon}. Linear Layouts provide an
algebraic representation for such compute-side layouts, mapping
hardware-resource coordinates to logical tensor coordinates over \fTwo
\cite{linearlayouts2025}. TileDRAM extends this abstraction to DRAM placement coordinates, making tile-to-DRAM placement a
compiler-visible layout decision.

\paragraph{Controller Mapping and Software-Directed DRAM Placement.}
Permutation-based interleaving distributes pages across DRAM structures
\cite{zhang2000permutation}; Liu et al.~\cite{liu2018valley} propose PAE, an
XOR-based mapping that improves bank/channel parallelism while preserving
row-buffer locality; SDAM exposes software-configurable mappings for a 3D
memory controller \cite{zhang2022sdam}; Rotaru et al.~synthesize an
application-specific mapping over \fTwo\ from a representative application
trace \cite{rotaru2025xordrammapping}; and FACIL selects a mapping for each
huge page \cite{seo2025facil}. These approaches treat the controller mapping as
the decision variable. A single workload-specific mapping must serve
heterogeneous accesses, while specializing different access patterns requires
multiple mapping contexts and controller-side selection support.
Software-directed placement instead chooses where data reside, at page
\cite{agarwal2015pageplacement}, DNN-layer \cite{pandey2024neurotap}, region
\cite{khadirsharbiyani2022dataconvection}, or stream
\cite{li2024streamndp} granularity; the stream-based design also
controls element order and cache placement in a distributed DRAM cache.
TileDRAM keeps the controller mapping fixed and instead specializes data
placement using recurring ordered tile-local access behavior. Many tile access
signatures can therefore coexist under one mapping, with each
placement realized through ordinary address arithmetic and without
controller or ISA changes.

\section{Conclusion and Future Work}
\label{sec:conclusion}

TileDRAM shows that DRAM placement can be an explicit compiler-controlled tile
layout decision while the memory-controller address mapping remains fixed.
Building on recurring tile-local access behavior as a stable specialization
unit, it extends GPU Linear Layouts to represent reusable tile-to-DRAM
placements selected per tile access signature. The resulting
compiler--runtime toolchain realizes these layouts through ordinary address
arithmetic, without new ISA support or remapping hardware.
Across 108 full-trace MoE FFN cases, TileDRAM reduces the request-weighted
row-conflict rate from 75.7\% to 15.2\% and achieves a 1.77$\times$ geomean
speedup; retargeting across six 3D DRAM organizations yields a 2.20$\times$
geomean speedup, while address transformation adds 0.17\% overhead on
an RTX~3090.

Our evaluation focuses on tiled GEMM within one CU--stack pair, with the 3D DRAM
memory system evaluated through trace-driven Accel-Sim/Ramulator2 replay.
Future work includes extending TileDRAM to other kernels, such as attention
and convolution, coordinating layouts
across multiple CU--stack pairs, and validating compiler-controlled placement on
physical 3D DRAM systems.


\bibliographystyle{ACM-Reference-Format}
\bibliography{references}

\clearpage
\appendix
\section{Tile-to-DRAM Layout Details}
\label{app:dram-layout-details}

This appendix formalizes the source interface for tile access, the semantic
tile-to-DRAM Linear Layout in a fixed target contract, the layout's
resident tiles and their DRAM regions, the intrinsic ordered row-buffer trajectory and its
physical realization, and the reconstruction of a realizable layout from its
projected row-reuse kernel. The projected reduction, fixed-domain solver
interface, kernel-to-layout reconstruction, and finite-catalog correctness are
developed in
Appendices~\ref{app:xor-weighted-diff}--\ref{app:xor-equivalence}.

Throughout this appendix, we fix one target contract and one
tile access signature $\theta$. Thus the access quantum, request-formation
policy, target mapping, selection parameters, and placement constraints are
standing assumptions rather than arguments repeated in every expression. We
write $\mathsf{dram}(\cdot)$ for the fixed target DRAM address mapping.

\subsection{Tile-to-DRAM Semantics}
\label{app:dram-domain-semantics}

\paragraph{Normalized access sequence.}
A signature $\theta=(\kappa,\rho,\chi)$ identifies the tile configuration
$\kappa$, the operand-property tuple $\rho$, and the ordered tile-local access
sequence $\chi$.
For one
representative access sequence of source addresses $(s_0,\ldots,s_{N-1})$ over interval
$[A,A+E)$, where $g$ divides $E$, access granularity $g$ gives

\[
  \begin{aligned}
    \ell_i&=\left\lfloor\frac{s_i-A}{g}\right\rfloor,
    &\delta_i&=(s_i-A)\bmod g,\\
    r_i&=\Lambda_\theta(\ell_i),
    &S&=(r_0,\ldots,r_{N-1}).
  \end{aligned}
\]

\begingroup
\sloppy

The fixed signature $\theta$, under the standing request-formation policy,
determines both the source interpretation and event order. Hence
$\Lambda_\theta$ maps each observed unit index directly
to its normalized coordinate, and $S$ is the only sequence consumed by layout
selection.

We require $\Lambda_\theta$ to be injective on the observed indices
$\{\ell_i\mid 0\le i<N\}$.

A different tile placement $j$ uses its own source base $A^{(j)}$, but every
admitted instance of the fixed signature yields the same sequence $S$.
Thus absolute addresses and runtime interleaving may change without changing
the input to offline layout selection associated with that tile access signature.
\par
\endgroup

\paragraph{Definition 1 (tile-to-DRAM Linear Layout).}
For $x\in\fTwo^n$, a reusable \emph{tile-to-DRAM Linear Layout} is the
block-structured map
\[
  W_{\mathrm{sem}}x
  =\begin{bmatrix}W_Cx\\W_Dx\\W_Rx\end{bmatrix},
  \qquad W_{\mathrm{sem}}\in\fTwo^{m\times n}.
\]
The three blocks select the residual address coordinate, row-buffer domain, and
row within that domain. If the candidate uses $b$ independent domain bits and
$r_{\mathrm{row}}$ additional row bits, then
$W_C\in\fTwo^{(m-b-r_{\mathrm{row}})\times n}$,
$W_D\in\fTwo^{b\times n}$, and
$W_R\in\fTwo^{r_{\mathrm{row}}\times n}$, with
$\operatorname{rank}(W_D)=b$ and
$\operatorname{rank}([W_D;W_R])=b+r_{\mathrm{row}}$. We assume $m\ge n$ and
$0\le b+r_{\mathrm{row}}\le n$, as required by the completion in
\appref{app:xor-reconstruction}. Every retained executable layout has full
column rank; \appref{app:xor-reconstruction} constructs such a completion.

Lemma~2 constructs one invertible target-basis matrix
$G\in\fTwo^{m\times m}$ whose domain and row blocks preserve the fixed mapper's
domain and row equality relations.

\begin{equation}
  W_{\mathrm{phys}}=G^{-1}W_{\mathrm{sem}}.
  \label{eq:semantic-to-physical-layout}
\end{equation}

We use $W_{\mathrm{sem}}$ directly for the reusable semantic layout.
For this layout, $\beta=(A,E,B)$ denotes one tile-placement record: source interval
$[A,A+E)$ and region base $B$. The fixed context and signature need not
be encoded explicitly in either object.

\paragraph{Proposition 1 (compiler composition closure).}
Let $P$ denote the compiler's composed Linear Layout from compute coordinates
to the normalized coordinates $r_i$. Then $W_{\mathrm{sem}}P$, its three output
blocks, and $W_{\mathrm{phys}}P$ are Linear Layouts. Distinct coordinates in
$\operatorname{im}(P)$ remain distinct exactly when
\[
  \ker(W_{\mathrm{sem}})\cap\operatorname{im}(P)=\{0\}.
\]
Full column rank of $W_{\mathrm{sem}}$ is sufficient.

\emph{Proof.}
Linear maps remain linear under composition and output slicing
\cite[Definition~4.2 and Proposition~4.8]{linearlayouts2025}; the invertible
basis change
$G^{-1}$ preserves rank and kernel. The
restriction is injective precisely when its image intersects the kernel only at
zero. \hfill$\square$

\paragraph{Canonical byte-address placement.}
The target supplies a fixed output translation $t_\rho$ and a byte-placement
function $\Pi_\rho$ whose outputs are $g$-aligned byte offsets. For each source
address $s\in[A,A+E)$, the bound layout computes

\[
\begin{gathered}
  \ell=\left\lfloor\dfrac{s-A}{g}\right\rfloor,\qquad
  \delta=(s-A)\bmod g,\qquad r=\Lambda_\theta(\ell),\\
  y=W_{\mathrm{phys}}r\oplus t_\rho,\qquad
  \mathsf{addr}(s;W_{\mathrm{sem}},\beta)=B+\Pi_\rho(y)+\delta.
\end{gathered}
\]

The within-unit byte offset $\delta$ is retained after layout placement. A legal
tile placement keeps it within the byte/column position of the same target domain and
row; $\Pi_\rho$ must also be in bounds and injective on the translated live
image. Request formation converts the resulting byte address into a request for
the fixed DRAM address mapping
$\mathsf{dram}$.

\paragraph{Lemma 1 (output-translation invariance).}
For a fixed output translation $t$,
\[
\begin{aligned}
  (W_{\mathrm{phys}}x\oplus t)\oplus
  (W_{\mathrm{phys}}x'\oplus t)
    &=W_{\mathrm{phys}}(x\oplus x'),\\
  W_{\mathrm{phys}}x\oplus t=W_{\mathrm{phys}}x'\oplus t
    &\Longleftrightarrow W_{\mathrm{phys}}x=W_{\mathrm{phys}}x'.
\end{aligned}
\]
Thus output translation preserves injectivity, pairwise differences, and
equality. Under the affine mapping-consistency conditions below, it only
relabels mapped domains and rows, so row-state classification remains a
property of $W_D$ and $W_R$.

\paragraph{Definition 2 (ordered row-buffer semantics).}

A bound instance is \emph{request-formation preserving} when it forms
one physical request $q_i$ for each normalized event, in the same source order
and with $r_i$ as its coordinate ancestry.
Physical addresses may vary across tile placements; the common ordered
sequence remains $S$. Define the semantic domain
and row labels
\[
  d_i=W_Dr_i,\qquad z_i=W_Rr_i.
\]
Immediately before request $i$, $\sigma_i(d)$ is the open row of domain $d$,
with $\bot$ denoting a closed domain. Starting from
$\sigma_0(d)=\bot$, request $i$ has class
\begin{equation}
  \epsilon_i=\begin{cases}
    \mathrm{miss},&\sigma_i(d_i)=\bot,\\
    \mathrm{hit},&\sigma_i(d_i)=z_i,\\
    \mathrm{conflict},&\text{otherwise},
  \end{cases}
  \qquad
  \sigma_{i+1}=\sigma_i[d_i\mapsto z_i].
  \label{eq:semantic-row-state-transition}
\end{equation}
Let

\[
  \mathsf{rowcnt}(S,W_{\mathrm{sem}})=
  \bigl(N_{\mathrm{conf}}(S,W_{\mathrm{sem}}),
        N_{\mathrm{miss}}(S,W_{\mathrm{sem}})\bigr).
\]

The two components count conflicts and opening misses in
Eq.~\eqref{eq:semantic-row-state-transition}; the remaining events are hits.
Full-trace replay applies controller scheduling, refresh, queueing, and timing
to the complete interleaved request sequence.
Thus, $\mathsf{rowcnt}$ characterizes the intrinsic source-order row-buffer
behavior of the semantic layout.

\paragraph{Definition 3 (mapping consistency).}

For physical request $q_i$, write
$\mathsf{dram}(q_i)=(d_i^{\mathrm{phys}},z_i^{\mathrm{phys}},\ldots)$, where
$d_i^{\mathrm{phys}}$ and $z_i^{\mathrm{phys}}$ are its resolved physical domain
and row.
A
bound instance is
\emph{mapping-consistent} when, for every pair of live requests,

\begin{align}
  d_i^{\mathrm{phys}}=d_j^{\mathrm{phys}}
    &\Longleftrightarrow d_i=d_j,\nonumber\\
  d_i=d_j&\Longrightarrow
  \bigl(z_i^{\mathrm{phys}}=z_j^{\mathrm{phys}}
    \Longleftrightarrow z_i=z_j\bigr).
  \label{eq:mapping-consistency}
\end{align}

The first equivalence says that $W_D$ names exactly the physical row buffers;
the second says that $W_R$ names rows within each such buffer. This pairwise
condition is checked on the finite admitted live image.

\paragraph{
Proposition 2 (semantic--mapped-label source-order equivalence).}

Let $\beta$ be a mapping-consistent, request-formation-preserving
tile placement of $W_{\mathrm{sem}}$.
Apply the transition in
Eq.~\eqref{eq:semantic-row-state-transition} to its mapped labels
$(d_i^{\mathrm{phys}},z_i^{\mathrm{phys}})$ and
call the resulting conflict and miss counts

$\mathsf{rowcnt}^{\mathrm{phys}}(S,W_{\mathrm{sem}},\beta)$. Then

\begin{equation}
  \mathsf{rowcnt}^{\mathrm{phys}}(S,W_{\mathrm{sem}},\beta)
  =\mathsf{rowcnt}(S,W_{\mathrm{sem}}).
  \label{eq:semantic-physical-rowcnt}
\end{equation}

The semantic and physical domain labels agree up to the bijective relabeling in
Definition~3.

\emph{Proof.}
Both executions start with all domains closed. By
Eq.~\eqref{eq:mapping-consistency}, they select corresponding domains and test
the same within-domain row equality at request $i$. They therefore emit the
same event and open corresponding rows. Induction proves identical event
sequences and preserves the realized domain labels up to relabeling.
\hfill$\square$

\paragraph{Lemma 2 (affine target factorization).}
Suppose the within-unit byte offset affects only the byte/column position, while byte
placement, request formation, and the fixed DRAM address mapping make the
physical row-buffer-domain and row labels affine in
$y=W_{\mathrm{phys}}x\oplus t_\rho$ on the admitted image. Let
$F_D\in\fTwo^{b\times m}$ and
$F_R\in\fTwo^{r_{\mathrm{row}}\times m}$ be their linear parts, with
$\operatorname{rank}(F_D)=b$ and
$\operatorname{rank}([F_D;F_R])=b+r_{\mathrm{row}}$.

The two rank conditions allow these row spaces to be extended to form an
invertible target-basis matrix $G$. Partition its rows according
to the residual, domain, and row outputs and require:

\[
  \begin{gathered}
    G=\begin{bmatrix}G_C\\G_D\\G_R\end{bmatrix}\in\fTwo^{m\times m},\\
    \operatorname{row}(G_D)=\operatorname{row}(F_D),\\
    \operatorname{row}\!\begin{pmatrix}G_D\\G_R\end{pmatrix}
    =\operatorname{row}\!\begin{pmatrix}F_D\\F_R\end{pmatrix}.
  \end{gathered}
\]

For $W_{\mathrm{phys}}$ from Eq.~\eqref{eq:semantic-to-physical-layout},
equality of physical row-buffer domains is exactly $W_D$ equality, and within
one domain physical row equality is exactly $W_R$ equality. Affine constants only relabel
these outputs. Definition~3 directly checks this correspondence on the admitted
live requests.

\emph{Proof.}

The two row-space equalities make $G_D$ preserve exactly the physical
domain-equality relation and make $[G_D;G_R]$ preserve exactly the joint
domain--row equality relation. Since
$GW_{\mathrm{phys}}=W_{\mathrm{sem}}$, these tests become equality of $W_Dx$
and, within a fixed domain, equality of $W_Rx$.
Hence $F_D$ equality is exactly $W_D$ equality, and, after fixing that domain
output, $F_R$ equality is exactly $W_R$ equality. Lemma~1 removes the affine
translation from both equality tests. \hfill$\square$

Thus same-$W_D$ predecessors are exactly predecessors in the same physical
row-buffer domain, and
$W_R$ equality within a domain is exactly physical row equality. The weighted
kernel therefore counts the
intrinsic source-order row hits of every certified realization, while
Proposition~2 preserves the source-order domain labels and row-buffer events.

\subsection{Source-Sequence Domain Coverage}
\label{app:domain-parallelism}

Source-sequence domain coverage ranks layouts after source-order row conflicts
have been minimized. It counts the independent row-buffer domains exercised by
the complete source sequence.

\paragraph{Definition 4 (source-sequence domain coverage).}
For semantic labels $d_i=W_Dr_i$ and ordered sequence
$S=(r_0,\ldots,r_{N-1})$, assume $N>0$.

Define the realized domain set and its coverage score as

\begin{equation}
  \begin{aligned}
    \mathcal D(S,W_D)&=\{W_Dr_i\mid 0\le i<N\},\\
    \mathsf{cov}(S,W_D)&=|\mathcal D(S,W_D)|.
  \end{aligned}
  \label{eq:domain-coverage}
\end{equation}

The score counts the distinct domains retained by the complete source-order
sequence. Under the closed-domain initial state of Definition~2, this same
distinct-domain count is also the opening-miss count established in
Appendix~\ref{app:xor-weighted-diff}. The selector therefore uses row conflicts
as its locality criterion and this coverage score as the next criterion. Let
\begin{equation}
  \mathcal C_{\mathrm{conf}}(S)
  =\operatorname*{arg\,min}_{W\in\mathcal W_{\mathrm{cert}}}
    N_{\mathrm{conf}}(S,W),
  \label{eq:domain-coverage-conflict-set}
\end{equation}
where the minimization ranges over the retained candidate family. Selection
then chooses
\begin{equation}
  W^\star\in
  \operatorname*{arg\,max}_{W\in\mathcal C_{\mathrm{conf}}(S)}
    \mathsf{cov}(S,W_D),
  \label{eq:domain-coverage-selection}
\end{equation}
before minimizing the executable map's nonzero-term complexity. Thus, among
layouts with identical conflict counts, the selected layout exercises the
largest number of distinct domains in the source sequence. The score depends only
on $(S,W_D)$, and Proposition~2
preserves it in every certified physical realization.

\subsection{Legal Placement and Allocation Invariance}
\label{app:materialization-legality}

A reusable layout must assign distinct live bytes to distinct backend addresses
and remain valid across the region bases admitted by the allocation policy.

\paragraph{Proposition 3 (address uniqueness).}
Let $W_{\mathrm{sem}}$ have full column rank. Suppose
$\Lambda_\theta$ is injective on the observed unit
indices, the region base $B$ and byte placement $\Pi_\rho$ are
$g$-aligned, and $\Pi_\rho$ is injective on the translated live image. If the
placement remains in bounds, then the byte-address function

$\mathsf{addr}(\cdot;W_{\mathrm{sem}},\beta)$
is injective over live layout units and their
within-unit byte offsets.

\emph{Proof.}
If two backend addresses are equal, alignment and reduction modulo $g$ give
$\delta=\delta'$. Cancelling the common base and offset, then using injectivity
of $\Pi_\rho$, gives
$W_{\mathrm{phys}}r\oplus t_\rho=
W_{\mathrm{phys}}r'\oplus t_\rho$. The translation cancels, and

$W_{\mathrm{phys}}=G^{-1}W_{\mathrm{sem}}$ has full column rank, so
$r=r'$.
Injectivity of
$\Lambda_\theta$ then gives the
same source-relative unit; together with $\delta=\delta'$, this gives the same
source byte.
\hfill$\square$

\paragraph{Definition 5 (legal allocation placement class).}
\begingroup
\sloppy

For a semantic layout $W$, let $\Gamma(W)$ be the set of tile placements
$\beta=(A,E,B)$ whose source interpretation and access quantum match the fixed
signature, whose byte placement satisfies Proposition~3 and remains inside the
reserved CU-local aperture, and whose physical mapping satisfies Definition~3.
DRAM regions of simultaneously resident tiles must be disjoint.
\par
\endgroup

\begingroup
\sloppy

A candidate family is request-formation stable when every
$\beta\in\Gamma(W)$ is request-formation preserving in the sense of
Definition~2, for every $W$ in the family.

Let $\mathcal W_{\mathrm{cert}}$ denote the finite certified catalog
retained by the declared synthesis policy. Every
$W\in\mathcal W_{\mathrm{cert}}$ has a nonempty set $\Gamma(W)$ and stable
request formation.
\par
\endgroup

\paragraph{Corollary 1 (allocation invariance).}
\begingroup
\sloppy

For any $W\in\mathcal W_{\mathrm{cert}}$ and any two tile placements
$\beta_1,\beta_2\in\Gamma(W)$,
\par
\endgroup

\[
  \begin{aligned}
    \mathsf{rowcnt}^{\mathrm{phys}}(S,W,\beta_1)
      &=\mathsf{rowcnt}(S,W)\\
      &=\mathsf{rowcnt}^{\mathrm{phys}}(S,W,\beta_2).
  \end{aligned}
\]
The source-sequence domain-coverage score also agrees across the class.

\emph{Proof.}
Apply Proposition~2 to each realization.
\hfill$\square$

\paragraph{Proposition 4 (periodic binding criterion).}

Suppose $T_{\mathrm{map}}$ is a positive multiple of $g$ and the fixed
target mapping repeats every $T_{\mathrm{map}}$ bytes up to one-to-one relabeling
of domains and rows. If $\beta=(A,E,B)\in\Gamma(W)$, then replacing $B$ by any
base in the same residue class modulo $T_{\mathrm{map}}$ yields another tile placement
in $\Gamma(W)$, provided that the resulting range stays inside the CU-local aperture.
Placements of simultaneously resident tiles are legal when their reserved regions are also
disjoint.

Formally, ``up to relabeling'' requires that, for each admitted base
shift $kT_{\mathrm{map}}$, fixed bijections on the realized domain labels and on
the realized rows within each domain transform the resolved labels at address
$a$ into those at $a+kT_{\mathrm{map}}$ for every live address $a$.

\emph{Proof.}
Changing the base by a multiple of
$T_{\mathrm{map}}$ preserves $g$-alignment and only relabels domains and rows, preserving their equality relations and hence
Definition~3. The shift is bijective, so address uniqueness is preserved; the
aperture and disjoint-range conditions supply the remaining placement
requirements.
\hfill$\square$

\emph{Instantiation.} The certified period of each admitted bit-field address
mapping is one row-block span. Starting from the base residue used during
synthesis, aligning each allocation extent to
$T_{\mathrm{map}}$ preserves that
residue for all contiguously packed allocations while keeping their ranges
disjoint.
Per-organization values of
$T_{\mathrm{map}}$ are recorded in \appref{app:artifact-configuration}.

\subsection{Projected Row-Reuse Kernel Reduction}
\label{app:xor-weighted-diff}

For the representative source-coordinate sequence
$S$ defined above,
let $W_D$ and $W_R$ represent the domain and row sublayouts of Definition~1.
Assume $\operatorname{rank}(W_D)=b$. Once $W_D$ is fixed, each domain's first
request is fixed; the remaining choice is whether later same-domain requests
reuse its open row.
The sequence partitions into domain-local subsequences. For every
non-first request, $\operatorname{pred}_D(i)$ denotes the preceding request in
the same $W_D$ domain.
Under
Definition~2's closed initial state, the intrinsic source-order opening-miss
count is

\begin{equation}
  N_{\mathrm{miss}}(S;W_D)
  =\left|\{W_Dr_i\mid 0\le i<N\}\right|
  =\mathsf{cov}(S,W_D).
  \label{eq:fixed-domain-misses}
\end{equation}
This count is independent of $W_R$. For a fixed $W_D$, opening misses and
coverage are therefore fixed; only the hit/conflict split varies with $W_R$.

Because $i$ and $\operatorname{pred}_D(i)$ have the same domain label,
their difference obeys

\[
\begin{aligned}
  \Delta_i
    &=r_i\oplus r_{\operatorname{pred}_D(i)},\\
  W_D\Delta_i&=0,\\
  W_Rr_i=W_Rr_{\operatorname{pred}_D(i)}
    &\Longleftrightarrow \Delta_i\in\ker(W_R)\\
    &\Longleftrightarrow \Delta_i\in K,
      \qquad K=\ker\!\begin{bmatrix}W_D\\W_R\end{bmatrix}.
\end{aligned}
\]

Collecting equal differences produces

\[
  w_{W_D}(\Delta)=
  \left|\left\{i\,\middle|\,
    \operatorname{pred}_D(i)\text{ exists and }\Delta_i=\Delta
  \right\}\right|.
\]
Summing these weights over the kernel gives the intrinsic hit count

\begin{equation}
  N_{\mathrm{hit}}(S;W_D,K)
  =\sum_{\Delta\in K}w_{W_D}(\Delta).
  \label{eq:weighted-hit-identity}
\end{equation}

Since every request is a miss, a hit, or a conflict,

\begin{equation}
  N_{\mathrm{conf}}(S;W_D,K)
  =N-N_{\mathrm{miss}}(S;W_D)-N_{\mathrm{hit}}(S;W_D,K).
  \label{eq:event-partition}
\end{equation}

For any completed layout
$W=[W_C;W_D;W_R]$ inducing $K$, these partial-layout counts are exactly
$N_{\mathrm{miss}}(S,W)$ and $N_{\mathrm{conf}}(S,W)$ used by
$\mathsf{rowcnt}(S,W)$.
Equations~\eqref{eq:fixed-domain-misses}--\eqref{eq:event-partition} prove that,
for fixed $W_D$, minimizing conflicts is exactly equivalent to maximizing the
weighted kernel objective.

Thus the row block affects the fixed-domain score only through its
row-reuse kernel $K$; two row blocks inducing the same $K$ capture the same
weighted differences.
If the joint domain--row map has rank
$b+r_{\mathrm{row}}$ and $n=b+r_{\mathrm{row}}+c$, then $\dim K=c$, and the
fixed-domain problem is

\begin{equation}
  \max_{K\subseteq\ker(W_D),\ \dim K=c}
  \sum_{\Delta\in K}w_{W_D}(\Delta).
  \label{eq:fixed-domain-subspace}
\end{equation}

Equation~\eqref{eq:fixed-domain-subspace} is the candidate-conditioned
fixed-domain kernel objective used by TileDRAM. For each candidate, the target
field widths and legal-placement constraints fix the admissible $b$,
$r_{\mathrm{row}}$, and $c=n-b-r_{\mathrm{row}}$. The next subsection gives
the reduced-coordinate solver interface and its lifting correctness.

\begin{table}[b]
\centering
\caption{The two-domain realization of the running $2\times2$ stream written as the
predecessor-difference calculation used by the solver.}
\label{tab:redline-kernel-example}

\small
\setlength{\tabcolsep}{4pt}
\begin{tabular}{@{}cccccc@{}}
\toprule
event & $r_i$ & $W_Dr_i$ & $\operatorname{pred}_D(i)$ & $\Delta_i$ & intrinsic event \\
\midrule
$e_0$ & $000$ & $D_0$ & --    & --          & miss \\
$e_1$ & $001$ & $D_0$ & $e_0$ & $001\in K$ & hit \\
$e_2$ & $010$ & $D_1$ & --    & --          & miss \\
$e_3$ & $011$ & $D_1$ & $e_2$ & $001\in K$ & hit \\
\bottomrule
\end{tabular}

\end{table}
The first request to each domain is an opening miss. The two remaining
requests contribute the same difference $001$, so choosing
$K=\operatorname{span}\{001\}$ captures weight two and makes both
intrinsic source-order hits.

\subsection{Fixed-Domain Solver Interface}
\label{app:fixed-domain-solver}

The solver operates only on the weighted same-domain predecessor differences.

After merging equal differences, choose $N_D$ whose columns form a basis
of $\ker(W_D)$.  The solver represents each difference by its unique coordinate
$u$ and searches a $c$-dimensional reduced subspace $\widetilde K$:

\[
\begin{gathered}
  N_D\in\fTwo^{n\times(n-b)},
  \qquad \operatorname{im}(N_D)=\ker(W_D),\\
  \widetilde K\subseteq\fTwo^{n-b},
  \qquad \dim\widetilde K=c,\\
  K=N_D\widetilde K
    :=\{N_Du\mid u\in\widetilde K\},\\
  \sum_{u\in\widetilde K}w_{W_D}(N_Du)
    =\sum_{\Delta\in K}w_{W_D}(\Delta).
\end{gathered}
\]

The resulting weighted-subspace interface is the one studied by Rotaru
et al.~\cite[\S6.1--6.2 and \S7]{rotaru2025xordrammapping}. TileDRAM uses the
cited linearized Problem~3 formulation for the exact route and its greedy
construction for the greedy route. The exact route optimizes a fixed $W_D$ on
the complete positive-weight support; the greedy route is heuristic. If the
returned generators span fewer than $c$ dimensions, they are extended to an
algebraically feasible $c$-dimensional subspace before lifting. Nonnegative
weights ensure that this completion cannot reduce the objective. The exact
route retains the highest-scoring such completion, whereas the greedy route
applies its feasible completion rule.

\paragraph{Proposition 5 (coordinate reduction and lifting correctness).}

Assume $\operatorname{rank}(W_D)=b$ and choose $N_D$ as above.  The map
$\widetilde K\mapsto N_D\widetilde K$ is a bijection from the
$c$-dimensional subspaces of $\fTwo^{n-b}$ to the feasible subspaces in
Eq.~\eqref{eq:fixed-domain-subspace}, and it preserves their objective values.
Consequently, the cited exact solution lifts to a fixed-$W_D$ optimum, while a
greedy solution lifts to a feasible fixed-$W_D$ solution.

\emph{Proof.}
\begingroup
\emergencystretch=2em

Because $N_D$ has full column rank, $u\mapsto N_Du$ is a bijection onto
$\ker(W_D)$ and preserves subspace dimension.
The displayed objective identity then preserves objective order after
equal differences have been merged.
\par
\endgroup
\hfill$\square$

\subsection{Kernel-to-Layout Reconstruction}
\label{app:xor-reconstruction}

\paragraph{Proposition 6 (row-block reconstruction).}
Let $K\subseteq\ker(W_D)$ have dimension $c$, assume $W_D$ has full row rank
$b$, and set $r_{\mathrm{row}}=n-b-c$. Then there is a row block
$W_R\in\fTwo^{r_{\mathrm{row}}\times n}$ such that
\[
  \ker([W_D;W_R])=K.
\]

\emph{Proof.}
From $K\subseteq\ker(W_D)$, it follows that
$\operatorname{row}(W_D)\subseteq K^\perp$. Extend a basis of
$\operatorname{row}(W_D)$ to a basis of $K^\perp$ using
$r_{\mathrm{row}}$ additional vectors, and use those vectors as the rows of
$W_R$. Then $\operatorname{row}([W_D;W_R])=K^\perp$, whose orthogonal
complement is $K$. \hfill$\square$

To complete the semantic map, Proposition~6 provides a joint domain--row block
of rank $b+r_{\mathrm{row}}=n-c$. Choose $c$ residual-address rows that add the
remaining independent directions and place them in $W_C$. This gives
\[
  W_{\mathrm{sem}}=\begin{bmatrix}W_C\\W_D\\W_R\end{bmatrix},\qquad
  \operatorname{rank}_{\fTwo}(W_{\mathrm{sem}})=n.
\]
Basis extension guarantees this full-rank algebraic completion. For $m>n$, a
candidate may add $m-n$ rank-redundant residual rows; the finite catalog retains
only completions satisfying the target embedding and sparsity constraints.
Lemma~2 expresses each retained completion as the executable map
$W_{\mathrm{phys}}$.

\subsection{Certified-Catalog Selection}
\label{app:xor-equivalence}

\begingroup
\sloppy

Domain-block enumeration followed by kernel and residual-address
reconstruction generates layout candidates.
Target realization, mapping-consistency, and binding-legality checks
retain the single finite certified catalog $\mathcal W_{\mathrm{cert}}$ defined
above.
\par
\endgroup

The complete lexicographic selection functional is

\begin{equation}
  J(S,W)=\bigl(
  N_{\mathrm{conf}}(S,W),
  -\mathsf{cov}(S,W_D),
  \operatorname{nnz}(G^{-1}W)\bigr).
  \label{eq:formal-selection-objective}
\end{equation}

The first coordinate minimizes source-order row conflicts. The second favors
greater source-sequence domain coverage from Eq.~\eqref{eq:domain-coverage}, and
the last is a nonzero-term complexity proxy for the executable linear address
map.

\paragraph{
Proposition 7 (catalog-relative exactness).}
Under Definition~4's assumption $N>0$, let

$\mathcal W_{\mathrm{cert}}$ be nonempty. By its definition, every
member is certified, preserves request formation, and has a nonempty legal
tile-placement set.
Exhaustive comparison returning

\[
  W^\star\in
  \operatorname*{arg\,min}_{W\in\mathcal W_{\mathrm{cert}}} J(S,W)
\]
is exact for the lexicographic objective over the emitted catalog. With the
exact inner route on the complete positive-weight support, the kernel emitted
for each retained $W_D$ is also optimal for its fixed-domain algebraic problem.
With the greedy inner route, each emitted kernel is feasible and exhaustive
comparison remains exact over the resulting catalog.

\emph{Proof.}
Definition~2 supplies the common ordered sequence, and Proposition~2 makes each
certified realization agree on the conflict coordinate; Definition~4 and
mapping consistency preserve the source-sequence domain-coverage score. The final
coordinate is computed directly from the emitted physical map. Proposition~5 supplies the exact or
feasible fixed-domain result for the chosen inner route. Since the catalog is
finite and nonempty, exhaustive comparison attains the catalog's lexicographic
minimum.
\hfill$\square$

\subsection{Applicability Conditions}
\label{app:generality-conditions}

The construction and its source-order analysis apply to recurring tiled access
patterns whose admitted instances satisfy three conditions; tiled GEMM is the
evaluated instance in this paper.

\emph{(C1) Recurring ordered sequence.}

For every $W\in\mathcal W_{\mathrm{cert}}$, the source interpretation
yields the same coordinate sequence for all
$\beta\in\Gamma(W)$, and request formation
preserves the one-request-per-event correspondence, event order, and coordinate ancestry. This
makes the row-state trajectory well defined and amortizes the cost of one offline
selection across the recurring tile placements.

\emph{(C2) Linear live-coordinate representation.} The source interpretation
maps live layout units injectively into $\fTwo^n$, and the compiler expresses
the resulting coordinate projection as a Linear Layout. With a selected
full-column-rank tile-to-DRAM Linear Layout, distinct live units therefore remain
distinguishable under composition.

\emph{(C3) Certified target realization.} On the admitted live image, the
within-unit byte offset affects only byte/column position, while byte placement and
the fixed DRAM address mapping produce affine domain and row labels with the
rank conditions of Lemma~2 and the consistency of Definition~3. The modeled
 intrinsic source-order state transition
retains one open row per domain until a request to that
domain replaces it.

Together, these conditions justify the ordered row-buffer semantics and
fixed-domain kernel reduction. Given a nonempty finite certified catalog, the
selector is well-defined. For an emitted $c$-dimensional kernel, its captured
same-domain predecessor-difference weight is exactly the source-order row reuse
from Eq.~\eqref{eq:weighted-hit-identity}.

\section{Controller-Side DRAM Address Mapping: Configuration Feasibility,
Construction Scale, and Performance}
\label{app:optimization-complexity}

Controller-side address mapping supplies one global interpretation for every
request served by the target memory controller. We evaluate fixed global hashes
that apply the same static bit-mixing rule to every operand stream, regardless
of its operand properties or tile access signature.
TileDRAM retains the target controller mapping and selects a tile-to-DRAM layout
for each tile access signature.

The experiments below separate construction cost from mapping effectiveness.
Section~\ref{app:rotaru-exhaustive-cost} evaluates the Rotaru-style exhaustive
full-trace search, whose outer mapping choice makes trace construction
candidate-dependent.

\subsection{Fixed Global Mappings: Closure and Replay Feasibility}

The fixed-hash family covers row-to-bank, column-to-bank, bank-to-row, and
column-to-row XOR mappings, compound row/bank swizzling, and a field swizzle
styled after Ramulator2's \texttt{MOP4CLXOR} address mapper
\cite{luo2023ramulator2}. Every recipe applies one frozen rule to
both operand access sequences. Compilation checks whether the rule closes over the live
address domain; replay-complete entries then execute through the same
Accel-Sim/Ramulator2 path as conventional placement and TileDRAM. The fixed-hash
replay uses a Llama~4 up/gate projection with tile
$(BM,BN,BK)=(64,128,64)$ under the \texttt{t8/b16/row13/col3} organization
with open-row FR-FCFS. The conventional
branch uses the identity \texttt{RoBaTiCoCh} mapping
(Row--Bank--Tier--Column--Channel), realized by Ramulator2's
\texttt{RoBaRaCoCh} mapper.

\begin{table}[t]
\centering
\scriptsize
\setlength{\tabcolsep}{1.5pt}
\caption{Feasibility and replay results for fixed global controller mappings on
a Llama~4 up/gate projection with tile $(64,128,64)$. Replay entries use the
refresh-on \texttt{t8/b16/row13/col3}
organization and report cycle-level Ramulator2 results relative to identity.
Higher conflict reduction and speedup are better.}
\label{tab:controller-fixed-hash-evidence}
\begin{tabularx}{\columnwidth}{@{}L{0.19\columnwidth}L{0.24\columnwidth}L{0.17\columnwidth}Y@{}}
\toprule
Mapping pattern & Address mixing & Validation & Recorded evidence \\
\midrule
Bank $\leftarrow$ Bank $\oplus$ LowRow
  & Low row bits into bank bits
  & Replay
  & 0.0000\% conflict reduction; 1.0000$\times$. \\
Bank $\leftarrow$ Bank $\oplus$ MidRow
  & Mid row bits into bank bits
  & Replay
  & 0.0000\% conflict reduction; 1.0000$\times$. \\
Bank $\leftarrow$ Bank $\oplus$ LowCol
  & Low column bits into bank bits
  & Replay
  & 0.0892\% conflict reduction; 1.0001$\times$. \\
Bank $\leftarrow$ Bank $\oplus$ LowRow $\oplus$ LowCol
  & Low row and column bits into bank bits
  & Replay
  & 0.0892\% conflict reduction; 1.0001$\times$. \\
Row $\leftarrow$ Row $\oplus$ LowBank
  & Low bank bits into row bits
  & Replay
  & 0.0000\% conflict reduction; 1.0000$\times$. \\
Row $\leftarrow$ Row $\oplus$ LowCol
  & Low column bits into row bits
  & Replay
  & 0.0000\% conflict reduction; 1.0000$\times$. \\
Compound bank/row XOR
  & Column-to-bank plus bank-to-row mixing
  & Replay
  & 0.0892\% conflict reduction; 1.0001$\times$. \\
\texttt{MOP4CLXOR}-style
  & Column-driven mapped-field swizzle
  & Replay
  & 0.0026\% conflict reduction; 1.0000$\times$. \\
\bottomrule
\end{tabularx}
\end{table}

The best evaluated replay-complete fixed global recipes reduce conflicts by only
0.0892\% and improve performance by 0.0144\%. The fixed mappings encode one
address-field transformation for the complete access sequence, whereas TileDRAM selects
placement from the ordered accesses of each signature.

\subsection{Cost of Rotaru-Style Exhaustive Full-Trace Synthesis}
\label{app:rotaru-exhaustive-cost}

The multi-bank mapping synthesis of Rotaru et al.~\cite{rotaru2025xordrammapping}
uses a complete workload trace as its specialization unit and treats the domain
mapping as an outer search variable. After fixing an outer mapping, the trace is
compressed into weighted same-domain predecessor differences for the
fixed-domain Problem~3 solver. Changing the outer mapping changes
request-domain membership, same-domain predecessors, and therefore the resulting
difference histogram. Trace construction is thus candidate-dependent: a separate
fixed-domain Problem~3 instance must be reconstructed for every outer candidate
before its inner subspace solve.

For the evaluated organization, the source coordinate spans $n=23$ bits and the
target requires $b=7$ tier-and-bank dimensions, corresponding to 128
row-buffer domains for a full-rank $W_D$. The bit-selecting family
then contains 245,157 candidates for $W_D$, while the structured-XOR family
contributes 105,417,617 raw candidates. Together they yield
105,662,774 outer candidates before rank filtering and row-space deduplication.
Applying this search directly to the Llama 4 full trace with 7,864,320
active-lane copy events would perform approximately $8.31\times10^{14}$
evaluations of candidate copy events before any
fixed-domain Problem~3 solve, full-map completion, or physical address-mapping
evaluation.

At a sustained aggregate rate of 10 billion evaluations of candidate copy events
per second on a large multicore CPU, this construction requires approximately
23.1 hours; at 1 billion evaluations per second, it requires approximately 9.6
days. These throughput-based estimates quantify raw outer construction before
Problem~3 solves, map completion, or physical address-mapping evaluation. The
outer construction alone therefore makes Rotaru-style exhaustive full-trace
controller-map synthesis computationally prohibitive as a routine
per-configuration path at the MoE FFN scale. Each new phase, problem shape, or
autotuned tile configuration supplies a new workload-wide specialization target.

These experiments separate synthesis tractability from mapping effectiveness.
Fixed global hashes are inexpensive but produce negligible changes in the
evaluated replay. Rotaru-style exhaustive full-trace synthesis explores a much
larger mapping space, but its candidate-dependent outer construction is
computationally prohibitive at this scale. TileDRAM instead changes the
specialization granularity: it keeps the controller mapping fixed and reuses
placement selected per recurring tile access signature.

\FloatBarrier
\Needspace{28\baselineskip}
\section{Compiler Interface}
\label{app:compiler-interface}

Listing~\ref{lst:tiledram-kernel-intent} shows an excerpt from the
programmer-facing tiled GEMM kernel used by the compiler path in
\secref{sec:tiledram-compiler-realization}. The annotation
exposes tiled-kernel semantics used by the compiler, while the kernel body uses
ordinary tiled loads and matrix multiplication; the output-store path is
omitted for brevity.

\noindent\begin{minipage}{\linewidth}
\begin{lstlisting}[style=tiledramcode,caption={Programmer-facing tiled GEMM
kernel excerpt used by the compiler path. The annotation exposes the tiled-kernel
semantics consumed by TileDRAM, while the body remains ordinary Triton/Gluon
code.},label={lst:tiledram-kernel-intent}]
@tiledram.gemm(
    problem=tiledram.problem(M=256, N=256, K=4096),
    tile=tiledram.tile(BM=64, BN=64, BK=64),
    execution=tiledram.execution(num_warps=4, num_stages=1),
    operands=tiledram.operands(
        A=tiledram.lhs("a_ptr", dtype="fp16",
                       layout="row_major", stride=("K", 1)),
        B=tiledram.rhs("b_ptr", dtype="fp16",
                       layout="row_major", stride=("N", 1)),
        C=tiledram.out("c_ptr", dtype="fp16",
                       layout="row_major", stride=("N", 1)),
    ),
)
@gluon.jit
def gemm_kernel(a_ptr, b_ptr, c_ptr, M, N, K,
                stride_am, stride_ak, stride_bk, stride_bn,
                stride_cm, stride_cn, BLOCK_M, BLOCK_N, BLOCK_K):
    a_ptrs = a_ptr + rows * stride_am + offs_k_a * stride_ak
    b_ptrs = b_ptr + offs_k_b * stride_bk + cols * stride_bn
    a_vals = ttgl.load(ttgl.set_auto_layout(a_ptrs, layout))
    b_vals = ttgl.load(ttgl.set_auto_layout(b_ptrs, layout))
    out_vals = ttgl.dot_fma(a_vals, b_vals, acc)
\end{lstlisting}
\end{minipage}

\paragraph{Tile-access and binding metadata.}
For each tile access signature, the compiler records the operand-property tuple,
comprising operand position (\texttt{lhs} or \texttt{rhs}), data type
(\texttt{dtype}), source-stride pattern (\texttt{stride}), and source layout
(row-major or column-major), together
with the access shape and address space needed by the lowered path, and routes
the resulting source pointer tensor through
\textsc{APPLY}. Compiler-side \textsc{BIND} performs template lookup and
records the selected target-specific layout template together with the rule for
deriving the tile ID from the current execution context. When that tile is
placed in a DRAM region, runtime \textsc{BIND} obtains the referenced template from this
binding metadata and combines it with the concrete tile ID, source interval, and
concrete DRAM region to construct its tile descriptor for that region.
Runtime \textsc{BIND} passes the descriptor for the active resident tile
as a runtime input to \textsc{APPLY}.

\paragraph{Address-lowering sequence.}
For each admitted operand access, lowering places the \textsc{APPLY} path after the
source pointer tensor is formed by address arithmetic. At execution, the path extracts the source-relative
coordinates, separates the layout-unit index from the within-unit byte
offset, applies the $\mathbb{F}_2$-linear map, affine XOR translation, and byte
placement, then reconstructs the backend address while retaining that offset.
Shifts, masks, XOR, parity, and address arithmetic implement this sequence
before the original asynchronous copy.

The compiler records the canonical address-function associated with each
transformed tile access signature. The compiler-side template reference and
runtime \textsc{BIND} connect the programmer-facing kernel intent, the
lowered compiler realization, and the simulator descriptor recorded by the
cross-path equivalence artifact.

\section{Artifact and Evaluation Configuration}
\label{app:artifact-configuration}

The artifact supports the three result groups reported in this paper: the
108-case MoE full suite and hardware cost calibration, the controller-mapping comparisons in
\appref{app:optimization-complexity}, and the six-organization retargeting
sweep. It records the revisions, build options, simulator
configuration files, trace manifests, and run manifests that determine the
reported results.

\paragraph{Software surface.}
The artifact pins the TileDRAM source tree, Accel-Sim/GPGPU-Sim build,
Ramulator2 checkout, Triton/Gluon compiler tree, CUDA and NVBit environment,
Python package set, host OS, C++ compiler, and simulator build flags. The run
manifests bind these fields to trace collection, layout selection, descriptor
construction, and replay for each result group. All reported TileDRAM
layouts are selected by the greedy route on the complete positive-weight support
(\appref{app:fixed-domain-solver}). Per-case synthesis-resource records support
the offline layout-synthesis cost reported below; controller-mapping manifests
support the comparisons in \appref{app:optimization-complexity}.

\paragraph{Cross-path equivalence artifact.}
The artifact includes a hash-closed compiler--simulator
execution-equivalence report. For every semantic address-function family, the
report records the representative descriptor and descriptor-table hashes,
canonical function hash, per-operand compiler transformation-binding and
lowered-IR hashes,
compiler binary hash, production transformer binary hash, structural-vector
hash, and mismatch count. The release gate passes only when the canonical
evaluator, lowered compiler IR, and production simulator transformer produce identical
addresses for every reported vector. The current report admits all 108
main-suite configurations and covers the identity and selected-layout
branches: 216 runtime descriptor tables (one per configuration per branch,
each covering both input operands), 2,118,464 descriptors
counted at per-tile-instance granularity, 378 semantic address-function
families, and 45,460 structural byte addresses with zero mismatches.

\paragraph{Compiler-cost calibration artifact.}
The hardware calibration uses the same 108 full-GEMM kernel configurations as
the main suite, spanning six model families and eight tile shapes. The selected
templates supply the address computation for both operands. Each case
records $H0$ without address transformation, $H1$ with identity address
generation, and $H2$ with the selected transformation; only $H2/H1-1$ enters
the reported overhead. Each mode contributes 51 measurements under a rotating
six-order
schedule and an adaptively sized timing window of at least 10~ms. The artifact
records the case, template, generated-code, binary, and
calibration-harness hashes together with the lowering report, resource report,
SASS dump,
and timing samples. The publication gate validates all 108 cases. The resulting
transformation-versus-identity overhead is 0.17\% geomean, below 0.01\% median,
0.57\% at the 95th percentile,
and 4.93\% maximum.

\paragraph{Offline layout-synthesis cost.}
TileDRAM performs one offline synthesis per case, producing the layouts for its
tile access signatures and amortizing the synthesis cost across their dynamic
tile instances. Across the
108-case solver-cost campaign, the measured median per-case synthesis time is
349.0~s and the measured median peak resident memory is 227.0~MB. The artifact
records the solver route, wall time, peak resident memory, and completion status
for each case.

\paragraph{Platform.}
GPU-side traces are collected from Triton and Gluon tiled GEMM kernels using
NVBit \cite{triton2019,tritonGluon,villa2019nvbit} and replayed in a trace-driven Accel-Sim/Ramulator2 platform. Accel-Sim supplies
the GPU execution context, including warp issue, instruction replay, memory
stalls, and cycle accounting. Ramulator2 supplies the 3D DRAM backend,
including address mapping, command scheduling, row-buffer state, read latency,
and foreground queue wait. \appref{app:optimization-complexity} separately
specifies the controller-mapping alternatives used in the controller-mapping
comparison.

\paragraph{Replay contract.}
Each baseline--TileDRAM pair is constructed from one CU-local access
sequence and one target DRAM organization. The paired fields are the kernel
shape, tile shape, execution staging, source trace, target organization, and
fixed DRAM address mapping; each branch records its observed source ingress and
controller service orders. The treatment
field is DRAM placement:
baseline uses conventional placement, while TileDRAM uses the selected
tile-to-DRAM layout for the same target organization.

\paragraph{Main workload set.}
The primary MoE full suite contains 108 full-trace FFN cases from
Qwen3 \cite{qwen3report2025}, DeepSeek-V3 \cite{deepseekv3report2024},
Qwen3.5-35B-A3B \cite{qwen3535ba3b2026}, Kimi K3 \cite{kimik3report2026},
Hunyuan-A13B \cite{hunyuana13b2025}, and Llama 4 \cite{llama4herd2025}, with
18 cases per family. Each family contributes one up/gate and one down
projection, and each projection is measured at $M=64$ over the four tiles
$64{\times}64{\times}32$, $64{\times}128{\times}32$, $64{\times}128{\times}64$,
$64{\times}256{\times}32$ and at $M=128$ over the five tiles
$64{\times}64{\times}32$, $128{\times}128{\times}32$,
$128{\times}128{\times}64$, $128{\times}256{\times}64$,
$128{\times}256{\times}128$, giving eight $M=64$ and ten $M=128$ cases per
family over eight distinct tile shapes; $64{\times}64{\times}32$ is the one
shape shared by both $M$ groups, so it carries 24 of the 108 cases while each
other shape carries 12. The suite also supplies the complete
hardware address-transformation cost cohort.
The artifact records the problem size, tile size, operand properties, including
operand position, data type, source-stride pattern, and source layout, and
model-family identifiers for each case. The retargeting sweep uses
six full-trace FFN streams with shapes
$(64,4096,1536)$, $(64,4096,7168)$, $(64,4096,3072)$,
$(128,8192,5120)$, $(128,3584,3072)$, and $(128,512,2048)$,
all with $64{\times}64{\times}32$ tiles and one pipeline stage.

\paragraph{Memory-system configurations.}
All modeled organizations use \texttt{OpenRowPolicy} and FR-FCFS. The full
suite, fixed-hash supplement, and retargeting sweep use
\texttt{RoBaTiCoCh} (Row--Bank--Tier--Column--Channel), realized by
Ramulator2's \texttt{RoBaRaCoCh} mapper. The full
suite uses 8 tiers,
16 banks, 13 row bits, and 3 column bits. Refresh is enabled for every replay
in the 108-case full suite. The fixed-hash supplement uses the
same tier/bank counts with 13 row bits and 3 column bits, as stated in
\appref{app:optimization-complexity}.
For the main organization, the address-mapping byte period is one row-block span,
$T_{\mathrm{map}}=128\times512$~bytes${}={}$64~KiB; the
corresponding spans for the retargeting organizations follow from their own
domain counts and row sizes. In particular, the two row12/col4 organizations
use 1-KiB rows and periods of 32~KiB and 128~KiB; the run manifests record the
value used for each organization.
The evaluated operand access sequence is an ordered tile-local access sequence of active-lane copy events generated
by LDGSTS.128; operand-tile source-coordinate sequences are the optimization
inputs, and all copy event and row-event counts use this same event
granularity. Because the CU-local aperture path has neither cross-CU request
merging nor a coalescing stage
(\secref{sec:bg-3ddram-chip}), each active-lane LDGSTS.128 event is one
16-byte request at the stack boundary, and both replay arms share this
granularity.
\paragraph{Controller-mapping artifacts.}
The artifact records each fixed-hash operation, scope, closure result, compile
score, replay manifest, and the separate operand-scoped diagnostic entry for
\appref{app:optimization-complexity}.

\paragraph{Metric definitions.}
The request-weighted row-conflict rate is the pooled conflict count divided by
all pooled row events. Pooled row-conflict reduction uses the corresponding
summed baseline and TileDRAM conflict counts. For performance, each
matched case contributes one speedup,
\[
  \mathit{SU}_i = \frac{C_{\mathrm{baseline},i}}{C_{\mathrm{TileDRAM},i}},
  \qquad
  \mathit{SU}_{\mathrm{geo}} =
  \left(\prod_{i\in\mathcal S} \mathit{SU}_i\right)^{1/|\mathcal S|}.
\]
Speedup is reported as the geometric mean of per-case
baseline-to-TileDRAM cycle ratios. Replay reports simulator-backed
memory-system performance;
RQ4 separately reports the hardware-measured address-transformation cost
(\secref{sec:eval-rq4-projection-cost}).
For organization $o$, we also report the pooled-cycle speedup
\[
  \mathit{P}_o =
  \frac{\sum_{i\in\mathcal W_o} C_{\mathrm{baseline},i,o}}
       {\sum_{i\in\mathcal W_o} C_{\mathrm{TileDRAM},i,o}},
\]
where $\mathcal W_o$ contains the six workload streams evaluated under $o$.
This organization-level pooled ratio is distinct from the six-stream
geometric mean; the pooled-cycle range in RQ3 is the minimum and maximum of
$\mathit{P}_o$ across the six organizations.

\paragraph{Reported metrics.}
The replay records row hits, row misses, row conflicts, row-buffer hit rate,
row-conflict rate, Ramulator2 average read latency, foreground queue wait,
shader memory-stall cycles, and cycles. The formal opening-miss count is
a source-order, closed-domain quantity; replay row-miss counters are service-order
observables that can also reflect refresh and controller behavior. The 108-case
MoE full suite reports 75.7\% baseline and 15.2\% TileDRAM row-conflict rates,
an 80.0\% pooled reduction, and a 1.77$\times$ geomean speedup.
Across the same cases,
shader-memory stalls fall by 53.9\% and request-weighted Ramulator2 read
latency falls by 49.6\%.
The retargeting experiment reports 36 paired cases across six 3D DRAM
organizations, with pooled row-conflict reduction of 79.2\% and geomean
speedup of 2.20$\times$.

\end{document}